\documentclass[prd,showpacs,floatfix,twocolumn,nofootinbib,amsmath,amssymb,floatfix]{revtex4}
\usepackage{graphicx,color,dcolumn,booktabs,bm}
\usepackage{longtable,lscape}
\usepackage{txfonts}
\usepackage{overpic}
\usepackage{amssymb}
\usepackage{indentfirst}
\usepackage{epsfig}
\usepackage{feynmf}   
\usepackage{epstopdf}   
\usepackage{slashed}  
\usepackage{cases}
\usepackage{color}
\usepackage{multirow}
\usepackage{graphicx,color,dcolumn,booktabs,bm}

\graphicspath{{Figures/}} %

\graphicspath{{Figures/}} %
\usepackage[colorlinks, citecolor=blue,anchorcolor=red,menucolor=red, linkcolor=red,filecolor=red,runcolor=red,urlcolor=blue,frenchlinks=red]{hyperref}
\begin{document}
\title{Light axial vector mesons}
\author{Kan Chen$^{1,2}$}\email{chenk_10@lzu.edu.cn}
\author{Cheng-Qun Pang$^{1,2}$}\email{pangchq13@lzu.edu.cn}
\author{Xiang Liu$^{1,2}$\footnote{Corresponding author}}\email{xiangliu@lzu.edu.cn}
\author{Takayuki Matsuki$^{3,4,1}$}\email{matsuki@tokyo-kasei.ac.jp}
\affiliation{$^1$School of Physical Science and Technology, Lanzhou University,
Lanzhou 730000, China\\
$^2$Research Center for Hadron and CSR Physics,
Lanzhou University $\&$ Institute of Modern Physics of CAS,
Lanzhou 730000, China\\
$^3$Tokyo Kasei University, 1-18-1 Kaga, Itabashi, Tokyo 173-8602, Japan\\
$^4$Theoretical Research Division, Nishina Center, RIKEN, Saitama 351-0198, Japan}

\begin{abstract}
Inspired by the abundant experimental observation of axial-vector states, we study whether the observed axial-vector states can be categorized into the
conventional axial-vector meson family. In this paper we carry out an analysis based on the mass spectra and two-body Okubo-Zweig-Iizuka-allowed decays.
Besides testing the possible axial-vector meson assignments, we also predict abundant information for their decays and the properties of some missing axial-vector mesons, which are valuable for further experimental exploration of the observed and predicted axial-vector mesons.
\end{abstract}

 \maketitle

\section{introduction}\label{sec1}

Among the light unflavored mesons listed in the Particle Data Group (PDG) \cite{Beringer:1900zz}, there are abundant light axial-vector mesons with a
spin-parity quantum number $J^{P}=1^{+}$, which form a $P$-wave meson family. Usually, we adopt $h_1$, $b_1$, $f_1$, and $a_1$ to express the corresponding states with the quantum numbers $I^G(J^{PC})=0^-(1^{+-})$, $1^+(1^{+-})$, $0^+(1^{++})$, and $1^-(1^{++})$, respectively. In Table \ref{meson list}, we collect the experimental information on the observed $h_1$, $b_1$, $f_1$, and $a_1$ states, as well as the corresponding resonance parameters and the observed decay channels.

Facing so many axial-vector states in the PDG, we need to examine whether all of these states can be categorized into the axial-vector meson family, which is
crucial for revealing their underlying structures. {We also notice that most axial-vector states are either omitted by the PDG or are recent findings needing confirmation. Due to the unclear experimental status of light axial-vector states, we
need to carry out a quantitative investigation of them, which would be helpful for further experimental studies, especially of those axial-vector states either omitted by the PDG or unconfirmed by other experiments.  }

In this work, we carry out a systematic study of the axial-vector states by analyzing mass spectra and
Okubo-Zweig-Iizuka (OZI)-allowed two-body strong decay behaviors. Our investigations
are based on the assumption that all of the axial mesons can be explained
within the conventional $q \bar q$ picture. Comparing our numerical results with the experimental data, we can further test the
possible assignments of the states in the axial-vector meson family. In addition, information on the predicted decays of the axial-vector states observed or still missing in experiments is valuable to further experimental study of axial-vector meson.

This paper is organized as follows. In Sec. \ref{sec2}, we present the phenomenological analysis by combining our theoretical results with the corresponding experimental data; the Regge trajectory analysis is adopted to study mass spectra of the axial-vector meson family and the quark-pair creation (QPC) model is applied to calculate their OZI-allowed strong decay behavior. Finally, the discussion and conclusion are given in Sec. \ref{sec3}.

\begin{table*}[htbp]
\caption{Resonance parameters and strong decay channels of the axial-vector states collected in the PDG \cite{Beringer:1900zz}. The masses and widths are average values taken from the PDG. The
states omitted from the PDG summary table are marked by a superscript $\natural$, while the states listed as further states in the PDG are marked by a superscript
$\flat$. \label{meson list}}
  \renewcommand{\arraystretch}{1.5}
\begin{tabular}{ccccccccc}
\toprule[1pt]
 $I^G(J^{PC})$&{State} &{Mass (MeV)}&{Width (MeV)} &The observed decay channels&
 \\     
 \midrule[1pt]
&$a_{1}(1260)$&        {$1230\pm40$} &    {$250\sim600$}   & $3\pi$ \cite{Barberis:2001bs}, $\pi\rho$ \cite{Asner:1999kj}, $\sigma \pi$ \cite{Chung:2002pu}  &  \\

& {$a_{1}(1640)^{\natural}$} &   {$1647\pm22$}  &{$254\pm27$}  &  $3\pi$ \cite{Baker:1999fc}, $\pi\rho$ \cite{Chung:2002pu,Amelin:1995gu} , $\sigma\pi$ \cite{Baker:1999fc}, $f_{2}(1270)\pi$ \cite{Baker:1999fc} \\     
$1^-(1^{++})$&$a_{1}(1930)^{\flat}$& $1930^{+30}_{-70}$ & $155\pm45$ & $3\pi^{0} $ \cite{Anisovich:2001pn}
\\
 &$a_{1}(2095)^{\flat}$ &  {$2096\pm17\pm121$}&  {$451\pm41\pm81$}& $\pi^{+}\pi^{-}\pi^{-}$ \cite{Kuhn:2004en}
 \\
& $a_{1}(2270)^{\flat}$  &$2270^{+55}_{-40}$   &$305^{+70}_{-40}$   &$3\pi^{0}$ \cite{Anisovich:2001pn}
 \\\midrule[1pt]
 &$b_{1}(1235)$  &  $1229.5\pm3.2$         &$142\pm9$ &$\omega\pi$ \cite{Karshon:1974qi,Amsler:1993pr,Nozar:2002br}
 \\
 $1^+(1^{+-})$&$b_{1}(1960)^{\flat}$   &$1960\pm35$ &$345\pm75$ &$\omega\pi^{0} $ \cite{Anisovich:2002su}
 \\
 &$b_{1}(2240)^{\flat}$   &$2240\pm35$  &$320\pm85$ &$\omega\pi^{0}$\cite{Anisovich:2002su}
 \\\midrule[1pt]
 &$f_{1}(1285)$   &$1282.1\pm0.6$   &$24.2\pm1.1$   &$\rho^{0}\rho^{0}$ \cite{Barberis:1999wn}, $\eta\pi\pi$ \cite{Dorofeev:2011zz,Corden:1978cz,Gurtu:1978yv}, $a_{0}\pi$ \cite{Corden:1978cz,Gurtu:1978yv,Bolton:1991nx}, $K\bar{K}\pi$ \cite{Barberis:1998by,Corden:1978cz,Gurtu:1978yv}
 \\
& $f_{1}(1420)$   &$1426.4\pm0.9$   &$54.9\pm2.6$   &$K \bar{K} \pi$ \cite{Bromberg:1980bk,Dionisi:1980hi}, $K \bar{K}^{*}(892)+$c.c \cite{Bromberg:1980bk,Dionisi:1980hi,Barberis:1998by}
 \\
$0^+(1^{++})$& $f_{1}(1510)^{\natural}$   &$1518\pm5$       &$73\pm25$     &$K \bar{K}^{*}(892)+$c.c \cite{Aston:1987ak,Birman:1988gu}, $\pi^{+}\pi^{-}\eta'$\cite{Ablikim:2010au}
 \\
& $f_{1}(1970)^{\flat}$   &$1971\pm15$ &$240\pm45$   &$\eta\pi^{0}\pi^{0}$ \cite{Anisovich:2000ut}
 \\
& $f_{1}(2310)^{\flat}$   &$2310\pm60$ &$255\pm70$   &$\eta\pi^{0}\pi^{0}$ \cite{Anisovich:2000ut}
 \\\midrule[1pt]
 &$h_{1}(1170)$          &$1170\pm20$  &$360\pm40$  &$\pi\rho$ \cite{Dankowych:1981ks,Atkinson:1983yx,Ando:1990ti}
 \\
& $h_{1}(1380)^{\natural}$&$1386\pm19$  &$91\pm30$   &$K\bar{K}^{*}(892)+$c.c \cite{Aston:1987ak,Abele:1997vu}
 \\
$0^-(1^{+-})$& $h_{1}(1595)^{\natural}$&$1594\pm15^{+10}_{-60}$  &$384\pm60^{+70}_{100}$   &$\omega\eta$ \cite{Eugenio:2000rf}
 \\
& $h_{1}(1965)^{\flat}$  &$1965\pm45$ &$345\pm75$  &$\omega\eta$ \cite{Anisovich:2011sva}
 \\
& $h_{1}(2215)^{\flat}$   &$2215\pm40$ &$325\pm55$ &$\omega\eta$ \cite{Anisovich:2011sva}
 \\
\bottomrule[1pt]
\end{tabular}
\end{table*}

\section{Phenomenological study of observed axial-vector states}\label{sec2}

A Regge trajectory analysis is an effective approach to study a meson spectrum \cite{Anisovich:2000kxa}, especially a light-meson spectrum. The masses and
radial quantum numbers of light mesons with the same quantum number
satisfy the following relation:
\begin{equation}
M^{2}=M_{0}^{2}+(n-1)\mu^{2},
\end{equation}
where $M_{0}$ and $M$ are the masses of the ground state and the corresponding radial excitation with radial quantum number $n$, respectively. $\mu^{2}$
denotes the slope of the trajectory with a universal $\mu^{2}=1.25\pm0.15$ GeV$^{2}$ \cite{Anisovich:2000kxa}.

\begin{figure*}[htbp]
\centering
\includegraphics[scale=0.65]{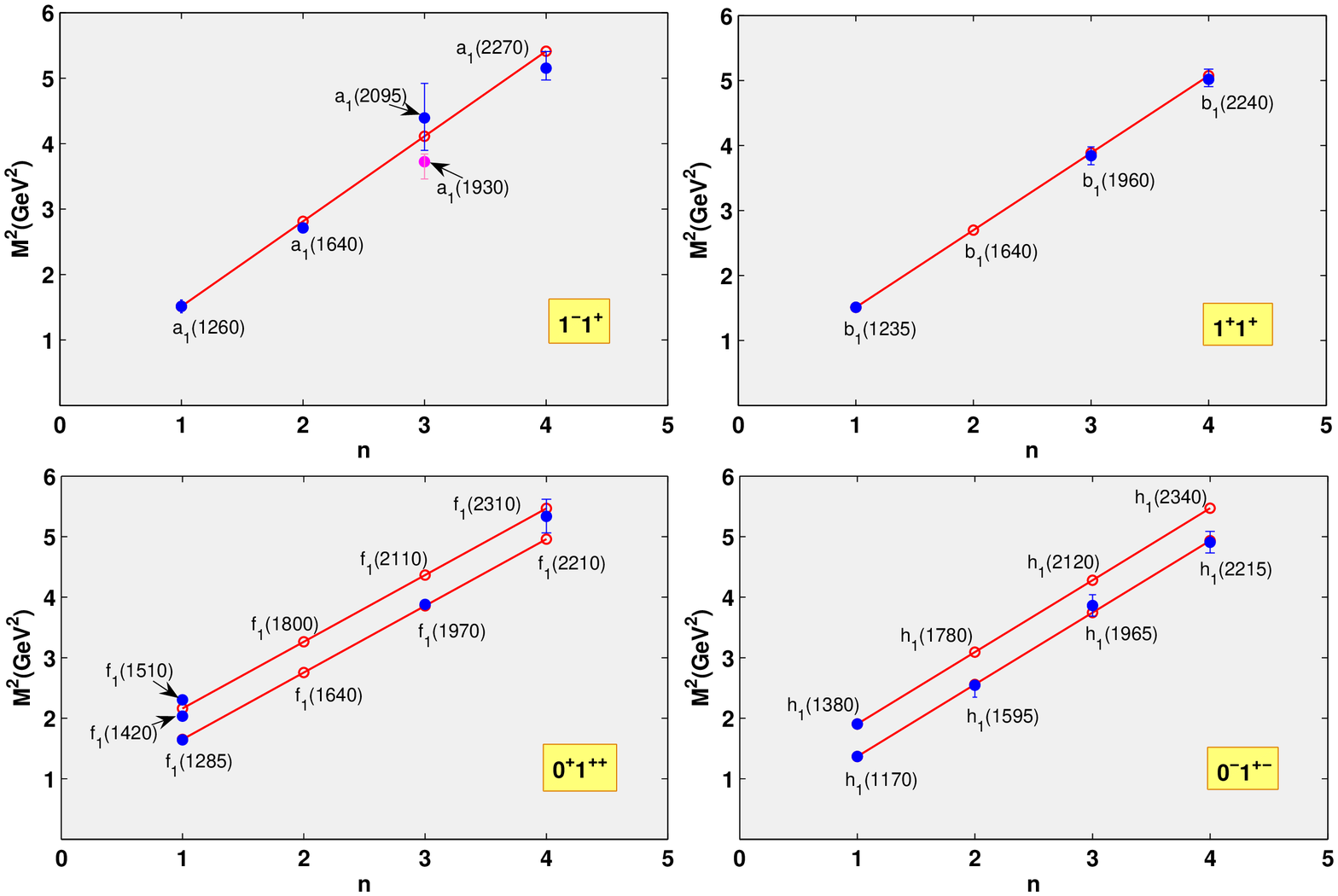}
\caption{(Color online) Regge trajectory analysis for $a_{1}$, $b_{1}$, $f_{1}$, and $h_{1}$ with typical $\mu^{2}=$ $1.30$ GeV$^{2}$, $1.19$ GeV$^{2}$, $1.10$
GeV$^{2}$, and $1.19$ GeV$^{2}$, respectively, {which can be covered by $\mu^2=1.25\pm0.15$ GeV$^2$ given in Ref. \cite{Anisovich:2000kxa}}.
The experimental errors of discussed axial-vector states are given, which are taken from the PDG \cite{Beringer:1900zz}.
Here, {{$\circ$}} and {\color{blue}{$\bullet$}} denote theoretical and experimental values,
respectively.
\label{Regge} }
\end{figure*}

In Fig. \ref{Regge}, we present the Regge trajectory analysis, in which we consider all of the axial-vector states listed in the PDG as shown in Table \ref{meson
list}. Besides the observed ones, we also predict some missing states and show them in Fig. \ref{Regge}. Additionally, we notice that there are two possible
candidates for the $a_1$ meson with quantum number $n^{2s+1}J_L=3^3P_1$, i.e., $a_1(1930)$ and $a_1(2095)$. On the other hand, both $f_1(1420)$ and $f_1(1510)$ can be an $s\bar{s}$ partner of $f_1(1285)$ by analyzing only the Regge trajectory. Thus, a further study of their strong decay behaviors
 would help to test these possible assignments to the observed axial-vector states and could provide more predictions of the observed and still-missing axial-vector mesons, which are valuable for the future experimental exploration of axial-vector mesons.

To obtain the decay behaviors of the axial-vector mesons, we adopt the QPC model, which was first proposed by Micu \cite{Micu:1968mk} and further developed by the Orsay group \cite{Le Yaouanc:1972ae,Le Yaouanc:1973xz,Le Yaouanc:1974mr,LeYaouanc:1977gm,Le Yaouanc:1977ux}. This model was
widely applied to study the OZI-allowed two-body strong decay of hadrons
\cite{vanBeveren:1979bd,vanBeveren:1982qb,Capstick:1993kb,Page:1995rh,Titov:1995si,Ackleh:1996yt,Blundell:1996as,
Bonnaz:2001aj,Zhou:2004mw,Lu:2006ry,Zhang:2006yj,Luo:2009wu,Sun:2009tg,Liu:2009fe,Sun:2010pg,Rijken:2010zza,Ye:2012gu,
Wang:2012wa,He:2013ttg,Sun:2013qca,Pang:2014laa,Wang:2014sea}. In the following, we briefly introduce
the QPC model.

For a decay process $A\to B+C$, we can write
\begin{eqnarray}
\langle BC|\mathcal{T}|A \rangle = \delta ^3(\mathbf{P}_B+\mathbf{P}_C)\mathcal{M}^{{M}_{J_{A}}M_{J_{B}}M_{J_{C}}},
\end{eqnarray}
where $\mathbf{P}_{B(C)}$ is a three-momentum of a meson $B(C)$ in the rest frame of a meson $A$. A subscript $M_{J_{i}}\, (i=A,B,C)$ denotes an orbital
magnetic momentum. The transition operator $\mathcal{T}$ is introduced to describe a quark-antiquark pair creation from vacuum, which has the quantum number
$J^{PC}=0^{++}$, i.e., $\mathcal{T}$ can be expressed as
\begin{eqnarray}
\mathcal{T}& = &-3\gamma \sum_{m}\langle 1m;1~-m|00\rangle\int d \mathbf{p}_3d\mathbf{p}_4\delta ^3 (\mathbf{p}_3+\mathbf{p}_4) \nonumber \\
 && \times \mathcal{Y}_{1m}\left(\frac{\textbf{p}_3-\mathbf{p}_4}{2}\right)\chi _{1,-m}^{34}\phi _{0}^{34}
\left(\omega_{0}^{34}\right)_{ij}b_{3i}^{\dag}(\mathbf{p}_3)d_{4j}^{\dag}(\mathbf{p}_4),
\end{eqnarray}
which is constructed in a completely phenomenological way to reflect the creation of a quark-antiquark pair from vacuum, where the quark and antiquark are
denoted by indices $3$ and $4$, respectively.
As a dimensionless parameter, $\gamma$ depicts the strength of the creation of $q\bar{q}$ from vacuum, where $\gamma=8.7$ and $8.7/\sqrt{3}$
\cite{Ye:2012gu} corresponds to the $u\bar{u}/d\bar{d}$ and $s\bar{s}$ creations, respectively. $\mathcal{Y}_{\ell m}(\mathbf{p})={|\mathbf{p}|^{\ell}}Y_{\ell
m}(\mathbf{p})$ is the solid harmonic. $\chi$, $\phi$, and $\omega$ denote the spin, flavor, and color wave functions, which can be treated separately. In
addition, $i$ and $j$ denote the color indices of a $q\bar{q}$ pair.

By the Jacob-Wick formula \cite{Jacob:1959at}, the decay amplitude is expressed as
\begin{eqnarray}
\mathcal{M}^{JL}(\mathbf{P})&=&\frac{\sqrt{4\pi(2L+1)}}{2J_A+1}\sum_{M_{J_B}M_{J_C}}\langle L0;JM_{J_A}|J_AM_{J_A}\rangle \nonumber \\
&&\times \langle J_BM_{J_B};J_CM_{J_C}|{J_A}M_{J_A}\rangle \mathcal{M}^{M_{J_{A}}M_{J_B}M_{J_C}},
\end{eqnarray}
and the general decay width reads
\begin{eqnarray}
\Gamma&=&\frac{\pi}{4} \frac{|\mathbf{P}|}{m_A^2}\sum_{J,L}|\mathcal{M}^{JL}(\mathbf{P})|^2,
\end{eqnarray}
where $m_{A}$ is the mass of an initial state $A$. We use the simple harmonic oscillator wave function to describe the space wave function of mesons,
which has the following expression:
\begin{eqnarray}
\Psi_{nlm}(R,\mathbf{p})= \mathcal{R}_{nl}(R,\mathbf{p}) \mathcal{Y}_{lm}(\mathbf{p}),
\end{eqnarray}
where the concrete values of the parameter $R$ involved in our calculation are given in Ref. \cite{Yu:2011ta} for the ground states. However, its value is to be fixed for each excited state.

With the above preparation, we further discuss the OZI-allowed decay behaviors of the axial-vector mesons. The
allowed decay modes are listed in Tables \ref{list1} and \ref{list2}.

\begin{table*}[htbp]
\begin{center}
 \renewcommand{\arraystretch}{1.3}
 \tabcolsep=0.65pt
\caption{OZI-allowed two-body decay channels for $a_{1}$ and $h_{1}$ states marked by $\surd$. Here, $\rho$, $\omega$, and $\eta$ denote $\rho(770)$,
$\omega(782)$, and $\eta(548)$, respectively. The axial-vector states predicted by the Regge trajectory analysis are marked by a superscript $\natural$.
\label{list1}}
\begin{tabular}{c|ccccc|c|cccccccc}
\toprule[1pt]
{Channel} & \footnotesize{$a_{1}(1260)$} & \footnotesize{$a_{1}(1640)$} & \footnotesize{$a_{1}(1930)$} &
\footnotesize{$a_{1}(2095)$}&\footnotesize{$a_{1}(2270)$} & {Channel} & \footnotesize{$h_{1}(1170)$} & \footnotesize{$h_{1}(1380)$}
&\footnotesize{$h_{1}(1595)$}& \footnotesize{$h_{1}^{\natural}(1780)$}& \footnotesize{$h_{1}(1965)$} & \footnotesize{$h_{1}^{\natural}(2120)$} &
\footnotesize{$h_{1}(2215)$} & \footnotesize{$h_{1}^{\natural}(2340)$} \\ 
 \midrule[1pt]
$\pi \rho$ & $\surd$ & $\surd$ & $\surd$ & $\surd$ & $\surd$& $\pi \rho$ &$\surd$ &$\surd$ &$\surd$ &$\surd$ &$\surd$&$\surd$&$\surd$ &$\surd$ \\
$\sigma \pi$    &$\surd$ & $\surd$ & $\surd$ & $\surd$ &    $\surd$&     $K K^{*}$  & &$\surd$ &$\surd$ &$\surd$ &$\surd$&$\surd$&$\surd$ &$\surd$      \\
$\pi f_{0}$ &$\surd$ & $\surd$ & $\surd$ & $\surd$ & $\surd$& $\eta \omega$ & &$\surd$ &$\surd$ &$\surd$&$\surd$&$\surd$&$\surd$ &$\surd$ \\
$ \pi f_{1}(1420) $ & & $\surd$ & $\surd$ & $\surd$ & $\surd$& $\omega\sigma$ && &$\surd$ & $\surd$ & $\surd$ & $\surd$ &$\surd$ &$\surd$ \\
$\pi \rho(1450) $     & & $\surd$ & $\surd$ & $\surd$ &    $ \surd$&    $K K_{1}(1270)$&   &&& $\surd$    &$\surd$ &$\surd$ &$\surd$ &$\surd$    \\
$\rho \omega$         & & $\surd$ & $\surd$ & $\surd$ &   $\surd$&       $\omega \eta'(958)$ & &&&$\surd$       &$\surd$ &$\surd$ &$\surd$   &$\surd$  \\
$\eta a_{0}(980)$     & & $\surd$ & $\surd$ & $\surd$ &   $\surd$&       $\omega f_{0}$       &        &&&$\surd$&$\surd$ &$\surd$ &$\surd$  &$\surd$     \\$K K^{*}$ & & $\surd$ & $\surd$ & $\surd$ & $\surd$& $\rho a_{0}(980)$ & &&&$\surd$&$\surd$ &$\surd$ &$\surd$ &$\surd$ \\
$\pi  b_{1}(1235)$          & & $\surd$ & $\surd$ & $\surd$ &  $\surd$& $\pi \rho(1450)$ &  &&      &$\surd$ &$\surd$ &$\surd$ &$\surd$ &$\surd$   \\

$\pi f_{2}(1270)$     & & $\surd$ & $\surd$ & $\surd$ &   $\surd$&       $K K_{1}(1400)$     &        &&&&$\surd$        &$\surd$ &$\surd$  &$\surd$    \\
 $\pi f_{1}(1285)$     & & $\surd$ & $\surd$ & $\surd$ &   $\surd$&      $K K^{*}(1410)$&        &        &&&$\surd$ &$\surd$ &$\surd$    &$\surd$  \\
$\rho a_{0}(980)$   & &         & $\surd$ & $\surd$ &     $\surd$&       $K K^{*}_{0}(1430)$ &        &  &&&$\surd$      &$\surd$ &$\surd$ &$\surd$    \\
$K K_{1}(1400)$      & &         & $\surd$ & $\surd$ &     $\surd$&     $K K^{*}_{2}(1430)$&        &  &&&$\surd$      &$\surd$ &$\surd$  &$\surd$   \\
$\eta a_{1}(1260)$        & &         & $\surd$ & $\surd$ &   $\surd$&      $K^{*} K^{*}$&        &   &&&   $\surd$  &$\surd$    &$\surd$&$\surd$    \\

$\pi \rho(1700)$      & &         & $\surd$ & $\surd$ &   $\surd$&          $\eta\omega(1420)$&        & &&&$\surd$       &$\surd$    &$\surd$&$\surd$  \\
$K K_{1}(1270)$ & & & $\surd$ & $\surd$ & $\surd$& $\sigma h_{1}(1170)$ & & && & $\surd$ &$\surd$ &$\surd$ &$\surd$\\
$K K^{*}(1410)$    & &         & $\surd$ & $\surd$ &   $\surd$&   $\pi \rho(1700)$        &    &&&   &$\surd$ &$\surd$ &$\surd$ &$\surd$ \\
$K K^{*}_{0}(1430)$         & &         & $\surd$ & $\surd$& $\surd$  &$\rho a_{2}(1320)$&       &        &   &&&$\surd$     &$\surd$ &$\surd$ \\
 $K K^{*}_{2}(1430)$     & &         & $\surd$ & $\surd$ & $\surd$&     $\omega f_{2}(1270)$&       &        &     &&&$\surd$   &$\surd$ &$\surd$    \\
$K^{*} K^{*}$       & &         & $\surd$  & $\surd$ & $\surd$&    $\sigma\omega(1420)$      &      &        & &&& $\surd$       &$\surd$ &$\surd$ \\
$\eta a_{2}(1320)$   & &         & $\surd$    & $\surd$& $\surd$ &$\omega f_{1}(1285)$ &   &&&    &        &   $\surd$     &$\surd$ &$\surd$  \\
$\sigma a_{1}(1260)$&&&$\surd$ &$\surd$ &$\surd$& $\rho\pi(1300)$&       &        & &&& $\surd$        &$\surd$ &$\surd$   \\
$\sigma a_{2}(1320)$&&&$\surd$ &$\surd$ &$\surd$&      $\rho a_{1}(1260)$&        & &&&       &$\surd$   &$\surd$    &$\surd$\\

$\rho \pi(1300)$  & &         &         & $\surd$ &  $\surd$&          $K^{*} K_{1}(1270)$&       & &&&       & &$\surd$&$\surd$ \\
$\eta a_{0}(1450)$      & &         &         & $\surd$ & $\surd$&  $f_{0}h_{1}(1170)$& &&&      &        & &$\surd$&$\surd$ \\

$\omega b_{1}(1235)$ &&         &         & $\surd$ &  $\surd$&                 $KK^{*}(1680)$&&&&       &        &        &$\surd$  &$\surd$\\

$\rho h_{1}(1170)$   & &         &   & $\surd$ &    $\surd$ & $\omega f_{1}(1420)$&  &&&     &        &        &$\surd$&$\surd$  \\
$\rho a_{1}(1260)$    & &         &         & $\surd$ & $\surd$ &$K^{*}K_{1}(1400)$    &      &        &&&&&&$\surd$    \\
$K^{*} K_{1}(1270)$   & &&         &         & $\surd$ &        &      &        & &\\
$\rho a_{2}(1320)$    & &&         &         & $\surd$        &&      &        &        &\\
$\rho \omega(1420)$   && &         &         & $\surd$                            &      &        &        &\\
$\rho a_{0}(1450)$&&&&&$\surd$&&&&&\\
$K K^{*}(1680)$     & &    &     &         & $\surd$&&&&&\\
$\eta'(958) a_{0}(980)$   & & &         &         & $\surd$&&&&&\\
 \bottomrule[1pt]
\end{tabular}
\end{center}
\end{table*}

\begin{table*}[htbp]
\begin{center}
 \renewcommand{\arraystretch}{1.3}
 \tabcolsep=0.65pt
\caption{OZI-allowed two-body decay channels for $b_{1}$ and $f_{1}$ states marked by $\surd$. Here, $\rho$, $\omega$, and $\eta$ denote $\rho(770)$,
$\omega(782)$, and $\eta(548)$, respectively. The axial-vector states predicted by the Regge trajectory analysis are marked by a superscript $\natural$.
\label{list2}}
\begin{tabular}{c|cccc|c|ccccccccc}
\toprule[1pt]
{Channel} & \footnotesize{$b_{1}(1235)$} & \footnotesize{$b_{1}^{\natural}(1640)$} & \footnotesize{$b_{1}(1960)$} & \footnotesize{$b_{1}(2240)$} & {Channel}
&\footnotesize{$f_{1}(1285)$} & \footnotesize{$f_{1}(1420)$} & \footnotesize{$f_{1}(1510)$} &\footnotesize{$f_{1}^{\natural}(1640)$}&
\footnotesize{$f_{1}^{\natural}(1800)$}&\footnotesize{$f_{1}(1970)$}& \footnotesize{$f_{1}^{\natural}(2110)$}& \footnotesize{$f_{1}^{\natural}(2210)$}&
\footnotesize{$f_{1}(2310)$} \\ 
 \midrule[1pt]
$\pi \omega$ & $\surd$ & $\surd$ & $\surd$ & $\surd$ &$\pi a_{0}(980)$ &$\surd$ &$\surd$ &$\surd$ &$\surd$ &$\surd$ &$\surd$ &$\surd$ &$\surd$ &$\surd$ \\
$\pi a_{0}(980)$ & $\surd$ & $\surd$ & $\surd$ & $\surd$ & $\sigma\eta$&$\surd$&$\surd$&$\surd$&$\surd$&$\surd$&$\surd$&$\surd$&$\surd$&$\surd$\\
$\pi a_{1}(1260)$ & & $\surd$ & $\surd$ & $\surd$ &$\pi a_{1}(1260)$ & &$\surd$ &$\surd$ &$\surd$ &$\surd$&$\surd$ &$\surd$ &$\surd$ &$\surd$ \\
$\pi a_{2}(1320)$ & & $\surd$ &$\surd$ & $\surd$ &$K K^{*}$ & &$\surd$ &$\surd$ &$\surd$ &$\surd$&$\surd$ &$\surd$ &$\surd$ &$\surd$ \\
$ \pi \omega(1420) $ & & $\surd$ & $\surd$ & $\surd$ &$\pi a_{2}(1320) $ & & &$\surd$ &$\surd$ &$\surd$&$\surd$ &$\surd$ &$\surd$ &$\surd$ \\
$\pi a_{0}(1450)$ & & $\surd$ & $\surd$ & $\surd$ &$\pi a_{0}(1450)$ & & & &$\surd$ &$\surd$&$\surd$ &$\surd$ &$\surd$ & $\surd$ \\
$\eta \rho$ & & $\surd$ & $\surd$ & $\surd$&$\eta f_{0}$ & & & &$\surd$ &$\surd$&$\surd$ &$\surd$ &$\surd$ &$\surd$ \\
$\rho \rho$ & & $\surd$ & $\surd$ & $\surd$ &$\rho \rho$ & & & &$\surd$&$\surd$&$\surd$ &$\surd$ &$\surd$ &$\surd$\\
$K K^{*}$ & &$\surd$ &$\surd$ & $\surd$&$\omega\omega$ & & & &$\surd$&$\surd$&$\surd$ &$\surd$ &$\surd$ &$\surd$\\
$\sigma\rho$              & & $\surd$           & $\surd$           &  $\surd$ &$\sigma\eta'$&&&&$\surd$&$\surd$&$\surd$&$\surd$&$\surd$&$\surd$\\
$\eta'(958) \rho$ & & & $\surd$ & $\surd$&$K K_{1}(1270)$ & & & &&$\surd$ &$\surd$ &$\surd$ &$\surd$ &$\surd$ \\
$\rho f_{0}$ & & &$\surd$ & $\surd$ &$ K^{*} K^{*}$ & & & & &$\surd$ &$\surd$ &$\surd$ &$\surd$ &$\surd$ \\
$\omega a_{0}(980)$ & & & $\surd$ & $\surd$ &$\omega h_{1}(1170) $ & & & &&&$\surd$ &$\surd$ &$\surd$ &$\surd$ \\
$K K_{1}(1270)$ & & & $\surd$ & $\surd$ &$\eta'(958) f_{0}$ & & & &&&$\surd$ &$\surd$ &$\surd$ &$\surd$ \\
$K K_{1}(1400)$ & & & $\surd$ &$\surd$ &$\eta f_{2}(1270)$ & & & &&&$\surd$ &$\surd$ &$\surd$ &$\surd$\\
$K K^{*}_{0}(1410)$ & & & $\surd$ & $\surd$ &$K K_{1}(1400)$ & & & &&&$\surd$ &$\surd$ &$\surd$ &$\surd$ \\
$K K^{*}_{0}(1430)$ & & &$\surd$ & $\surd$ &$K K^{*}(1410) $ & & & &&&$\surd$ &$\surd$ &$\surd$ &$\surd$ \\
$K K^{*}_{2}(1430)$ & & & $\surd$ & $\surd$ &$K K^{*}_{0}(1430)$ & & & &&&$\surd$ &$\surd$ &$\surd$ &$\surd$ \\
$\sigma b_{1}(1235)$ & & & $\surd$ & $\surd$ &$ K K^{*}_{2}(1430)$ & & &&&&$\surd$ &$\surd$ &$\surd$ &$\surd$ \\
$K^{*} K^{*}$ & & &$\surd$ & $\surd$ &$\eta f_{1}(1285) $ & & & && &$\surd$ &$\surd$ & $\surd$ &$\surd$ \\
$\omega a_{2}(1320)$    &         &         &         & $\surd$        & $\sigma f_{2}(1270)$&&&&&&$\surd$&$\surd$&$\surd$&$\surd$\\
$ \omega \pi(1300)$   &         &         &         & $\surd$     & $\sigma f_{1}(1285)$&&&&&&$\surd$&$\surd$&$\surd$&$\surd$\\
$K^{*} K_{1}(1270)$                      & &         &         & $\surd$                  &$\sigma f_{1}(1420)$&&&&&&&$\surd$&$\surd$&$\surd$\\
$\eta \rho(1450)$ & & & & $\surd$ & $ \rho b_{1}(1235) $ & & & & &&&$\surd$ &$\surd$ &$\surd$ \\

$K K^{*}(1680)$ & & & & $\surd$ &$\eta f_{1}(1420)$ & & & & &&&$\surd$&$\surd$&$\surd$ \\
$a_{0}(980)h_{1}(1170)$ & & & & $\surd$ &$\omega \omega(1420)$ & & & & &&&&$\surd$ &$\surd$ \\
$\rho f_{2}(1270)$& & & & $\surd$ & $K^{*} K_{1}(1270)$ & & & & &&&& $\surd$ &$\surd$ \\
$\rho f_{1}(1285)$&         &         &         & $\surd$       &  $K K^{*}(1680)$              &        &        &        &     &&&&  &$\surd$     \\
$\rho f_{1}(1420)$&         &         &         &$\surd$        &$f_{0}f_{2}(1270)$               &        &        &        &  &&&&      &$\surd$  \\

$\rho b_{1}(1235)$ & & & & $\surd$ &$f_{0}f_{1}(1285)$ & & & &&&&& &$\surd$ \\
$\omega a_{1}(1260)$ & & & &$\surd$ &$a_{0}(980)a_{1}(1260)$ & & & &&&& & &$\surd$ \\
 &&&&&$a_{0}(980)\pi(1300)$       &&&&&&&&&$\surd$\\
&&&&&$a_{0}(980)a_{2}(1320)$       &&&&&&&&&$\surd$\\
 &&&&&$\eta'(958) f_{2}(1270)$          &&&&&&&&&$\surd$\\
       &&&&&$\rho \rho(1450)$          &&&&      &&& & &$\surd$\\&&&&&$\eta'(958) f_{1}(1285)$            &&&&                      &&&& &$\surd$\\
                                         & &       &       &                 & $K^{*} K_{1}(1400)$&&&&           &&&&&$\surd$\\

\bottomrule[1pt]
\end{tabular}
\end{center}
\end{table*}

\subsection{$a_{1}$ states}

\begin{figure}[htb]
\centering
\includegraphics[scale=0.52]{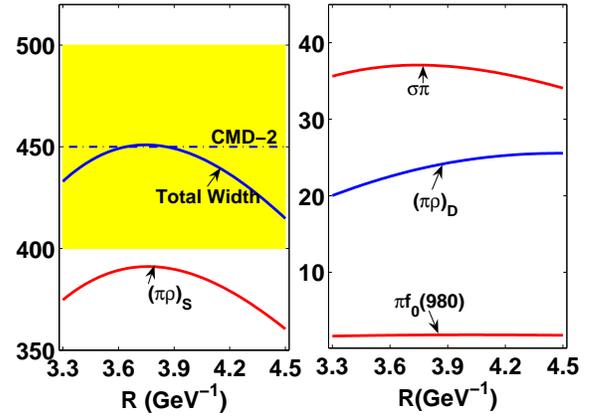}
\caption{(Color online) Total and partial decay widths of $a_{1}(1260)$ depending on $R$. Here, the dot-dashed line with the band is taken from the experimental data
from Ref. \cite{Akhmetshin:1998df}. The $S$-wave and $D$-wave contributions to the decay width of $a_{1}(1260)\to \pi\rho$ are also given separately. All results are in units of MeV.
\label{a11260} }
\end{figure}

The Regge trajectory analysis indicates that $a_{1}(1260)$ can be regarded as a ground state. The obtained total and partial decay widths of $a_{1}(1260)$
are listed in Fig. \ref{a11260}, which shows that $\pi\rho$ is the dominant channel. In Fig. \ref{a11260}, we give the partial decay widths of $a_{1}(1260)
\rightarrow \pi\rho$ from the $S$-wave and $D$-wave contributions.
Here, the $D$-wave/$S$-wave amplitude ratio in the decay $a_{1}(1260) \rightarrow \pi\rho$ is $-0.248$ with a typical value of $R=3.846$ GeV$^{-1}$
\cite{Yu:2011ta} in our calculation, which is comparable with the B852 data ($-0.14\pm0.04\pm0.07$) \cite{Chung:2002pu}. Our result also shows that
$a_{1}(1260) \rightarrow f_{0}\pi$ is a subordinate decay mode with the partial decay width 1.82 MeV, which explains why there has been no evidence of
$a_{1}(1260) \rightarrow f_{0}\pi$ in experiments \cite{Asner:1999kj}. As shown in Fig. \ref{a11260}, the calculated total width can reproduce the CMD2 data
given in Ref. \cite{Akhmetshin:1998df}.
In addition, we also give some typical ratios relevant to the partial decay and total widths together with the corresponding experimental data in Table \ref{a11260r}. In summary, our results are comparable with the experimental values and support $a_{1}(1260)$ as a ground state in the $a_1$ meson family.

\begin{table}[htbp]
 \renewcommand{\arraystretch}{1.2}
\caption{Some typical ratios of decay widths of $a_1(1260)$. The $\Gamma(\pi \rho)_{S(D)}$ represent the $S(D)$-wave decay width of $a_1(1260)\to \pi\rho$.
 \label{a11260r}}
\begin{tabular}{lcccccccc}
\toprule[1pt]
&                                                  {Our work} &                    {Experimental data}  \\     
 \midrule[1pt]
$\Gamma ((\pi \rho)_{S})/\Gamma _{Total}$ &                       $0.86$ &            0.60 \cite{Asner:1999kj} \\     
$\Gamma ((\pi \rho)_{D})/\Gamma _{Total}$ & $5.3\times10^{-2}$ & $(1.30\pm 0.60\pm0.22)\times10^{-2}$~ \cite{Asner:1999kj} \\ 
$\Gamma _{\pi\sigma}/\Gamma _{Total}$ &                    $8.2\times10^{-2}$       &$(18.76\pm4.29\pm1.48)\times10^{-2}$~ \cite{Asner:1999kj}  \\     
$\Gamma _{\sigma \pi}/\Gamma _{(\rho \pi)_{S}}$ &                          {$0.09$} &              {$0.06\pm0.05$}~\cite{Beringer:1900zz}   \\     
\bottomrule[1pt]
\end{tabular}
\end{table}

\begin{figure}[htbp]
\includegraphics[scale=0.42]{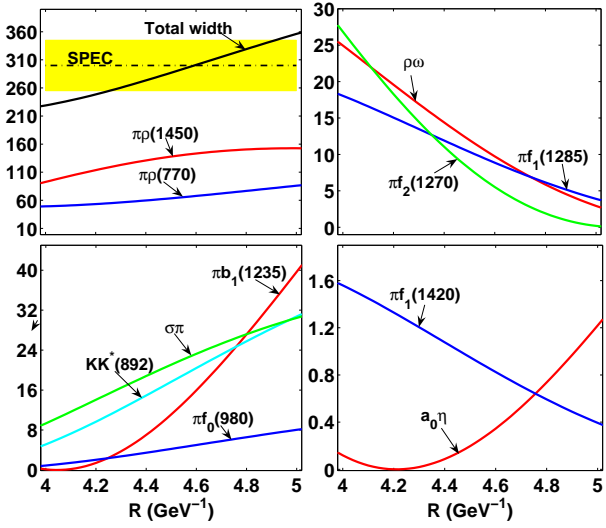}
\caption{(Color online) $R$ dependence of the decay behaviors of $a_{1}(1640)$. Here, the dot-dashed line with the band is the experimental total
width from Ref. \cite{Baker:1999fc}. All results are in units of MeV. \label{a11640}}
\end{figure}

If $a_{1}(1640)$ is the first radial excitation of $a_{1}(1260)$, its decay behavior depending on the $R$ value is shown in Fig. \ref{a11640}.
We use the experimental total width \cite{Baker:1999fc} and the ratio $\Gamma(f_{2}(1270)\pi)/\Gamma(\sigma\pi)=0.24\pm0.07$ \cite{Baker:1999fc} to get $R=(4.30\sim4.64)$ GeV$^{-1}$\footnote{Using the experimental total width \cite{Baker:1999fc}, we find that overlap exists between our theoretical and experimental results when taking $R=4.26\sim4.92$ GeV$^{-1}$. Then, we can further constrain the $R$ values by the ratio $\Gamma(f_{2}(1270)\pi)/\Gamma(\sigma\pi)=0.24\pm0.07$ \cite{Baker:1999fc}, where the constrained $R=(4.30\sim4.64)$ GeV$^{-1}$, which is adopted to present other typical ratios of
$a_{1}(1640)$. }. The main decay modes of $a_{1}(1640)$ are $\pi\rho$, $\pi\rho(1450)$, $\pi f_{2}(1270)$, $\pi f_{1}(1285)$, and $\rho\omega$.   Additionally, we provide further information on the typical ratios of $a_{1}(1640)$ decays in Table \ref{a11640r}.

\begin{table}[htbp]
 \renewcommand{\arraystretch}{1.2}
\caption{Typical ratios of the decay widths of $a_1(1640)$ corresponding to the $R$ range $(4.30\sim4.64)$ GeV$^{-1}$.
 \label{a11640r}}
\begin{tabular}{lcccccccc}
\toprule[1pt]
Ratio&Value&Ratio&Value\\
\midrule[1pt]
$\Gamma_{\pi\rho}/\Gamma_{Total}$&$0.216\sim0.227$&$\Gamma_{\pi\rho(1450)}/\Gamma_{Total}$&$0.473\sim0.474$\\
$\Gamma_{\pi b_{1}(1235)}/\Gamma_{Total}$&$0.014\sim0.059$&$\Gamma_{KK^{*}}/\Gamma_{\sigma\pi}$&$0.166\sim0.221$\\
$\Gamma_{\pi f_{2}(1270)}/\Gamma_{\rho\omega}$&$0.523\sim0.855$&$\Gamma_{\pi f_{1}(1420)}/\Gamma_{\pi f_{1}(1285)}$&$0.089\sim0.094$\\
$\Gamma_{\pi f_{0}}$&$0.166\sim0.221$\\

\bottomrule[1pt]
\end{tabular}
\end{table}

\begin{figure}[htbp]
\includegraphics[scale=0.38]{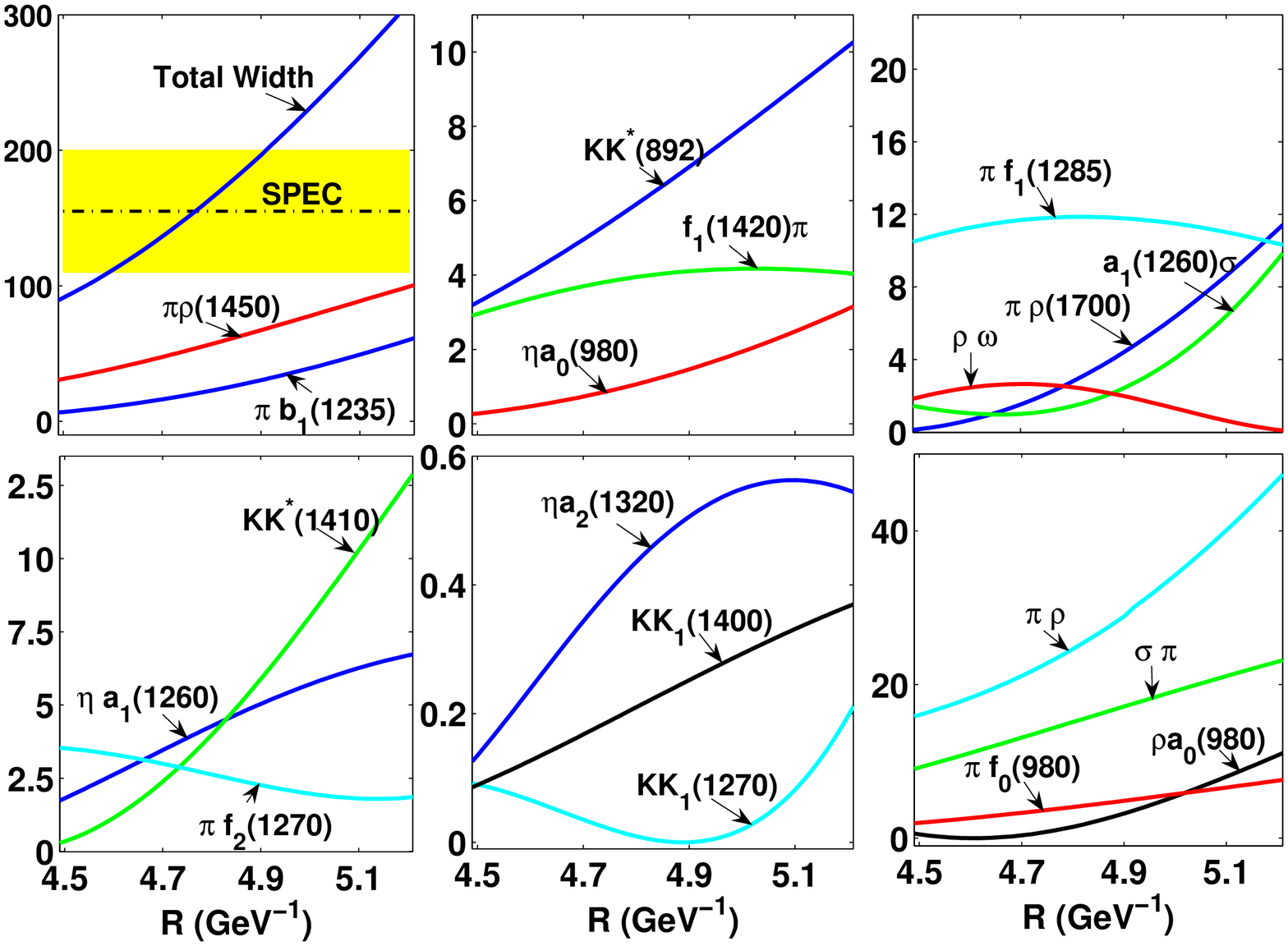}
\caption{(Color online) $R$ dependence of the calculated partial and total decay widths of the $a_{1}(1930)$. Here, the dot-dashed line with band is the experimental total width from Ref. \cite{Anisovich:2001pn}. All results are in units of MeV.
\label{a11930}}
\end{figure}

\begin{figure*}[htbp]
\includegraphics[scale=0.45]{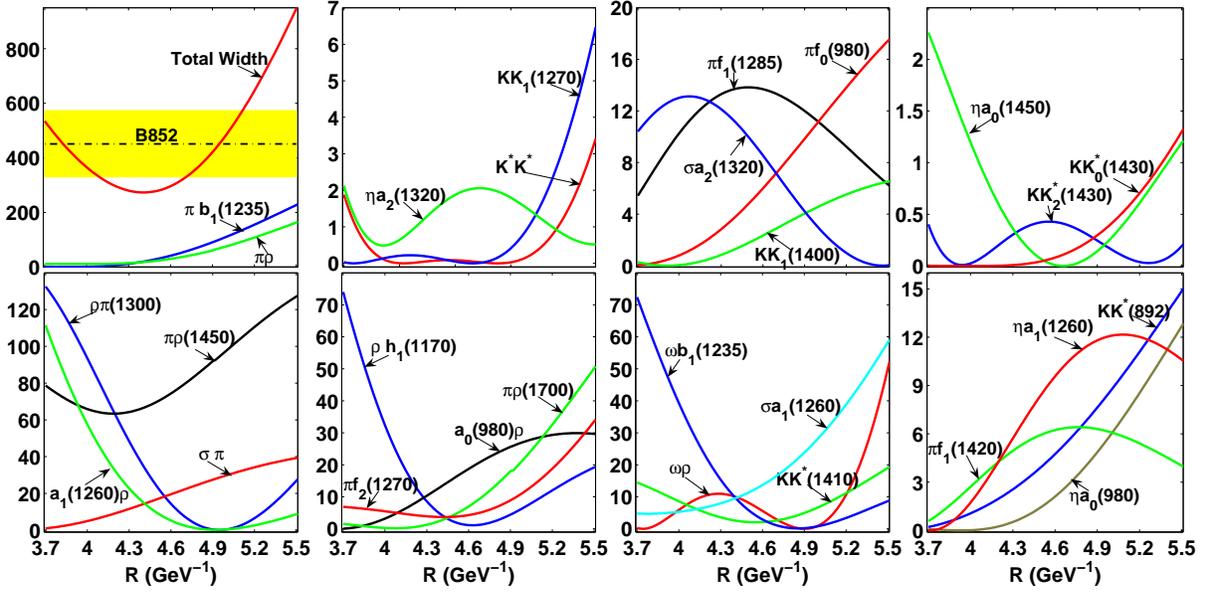}
\caption{(Color online) $R$ dependence of the calculated partial and total decay widths of $a_{1}(2095)$. Here, the dot-dashed line with the band is the
experimental total width from Ref. \cite{Anisovich:2001pn}. All results are in units of MeV. \label{a12095}}
\end{figure*}

There are two possible candidates for the second radial excitation of $a_1(1260)$. In the following, we discuss the decay behaviors of $a_{1}(1930)$ and $a_{1}(2095)$ by combining the corresponding experimental data. In Figs. \ref{a11930} and \ref{a12095}, we present the $R$ dependence
of the decay behaviors of these $a_1$'s, respectively. 
That is,
the obtained total width of $a_{1}(1930)$ can be fitted with the data from Ref. \cite{Anisovich:2001pn} when $R=4.58\sim 4.92$ GeV$^{-1}$, while that of $a_1(2095)$
can overlap with the experimental data \cite{Anisovich:2001pn} when $R=(4.78\sim5.16)$ GeV$^{-1}$.
Thus, it is difficult to distinguish which $a_1$ is more suitable as a candidate for the second radial excitation of $a_1(1260)$ by studying only the total decay widths. Besides, we can learn from the Regge trajectory analysis that there is only one state for the $3^{3}P_{1}$ state, and it is doubtful that both $a_{1}(1930)$ and $a_{1}(2095)$ exist, as mentioned in Ref. \cite{Anisovich:2001pn}.
However, there exist different behaviors of the partial decay widths of these $a_{1}$'s. The $a_{1}(1930)$ mainly decays into final states $\pi \rho$, $\pi \rho(1450)$, and $\pi b_{1}(1235)$, while the $\pi f_{1}(1285)$ and $\sigma\pi$ modes also have sizable contributions. The decays of $a_{1}(1930)$ into $K K^{*}_{0}(1430)$, $K K^{*}_{2}(1430)$, and $K^{*}(896)K^{*}(896)$ have tiny decay widths, which are not listed in Fig. \ref{a11930}. As for $a_{1}(2095)$, its dominant decay channels are $\pi b_{1}(1235)$, $\pi\rho$, and $\pi\rho(1450)$ and are shown in Fig. \ref{a12095}. The other decay channels---like $\rho
a_{0}(980)$, $\pi\rho(1700)$, $\pi f_{1}(1285)$, $\pi f_{0}$, and $\sigma\pi$---also have considerable contributions to the total decay width. In Table
\ref{a119302095r}, we also list some typical ratios relevant to their decays.
{We still need to emphasize one point. At present, $a_{1}(1930)$ and $a_{1}(2095)$ are not well established in experiments. The authors of Ref. \cite{Anisovich:2001pn} indicated that
{\it $a_2(1950)$ and $a_1(1930)$ are not securely identified in mass and width, though some such contributions are definitely required}  \cite{Anisovich:2001pn}. However, when considering the Regge trajectory analysis,  one finds that the $3^{3}P_{1}$ state in the $a_1$ meson family has a mass around 2000 MeV. The two unconfirmed $a_{1}(1930)$ and $a_{1}(2095)$ states could be candidates for the $3^{3}P_{1}$ state in the $a_1$ meson family, since their masses are close to that of the $3^{3}P_{1}$ state in the $a_1$ meson family. }
Thus, the experimental study of the partial decay widths of $a_{1}(1930)$
and $a_{1}(2095)$ will help to reduce the two possible candidates---$a_{1}(1930)$ and $a_{1}(2095)$---of the second radial excitation of the $a_1(1260)$ to one.
{In the following, the experimental confirmation of $a_{1}(1930)$
and $a_{1}(2095)$ will be crucial for identifying the candidate of the $3^{3}P_{1}$ state in the $a_1$ meson family.
If  $a_{1}(1930)$ and $a_{1}(2095)$ cannot be established in experiments, we suggest an experimental search for $a_1(3^{3}P_{1})$; the results for $a_1(3^{3}P_{1})$ predicted in this work
would be helpful for such a search. }

\begin{table}[htbp]
 \renewcommand{\arraystretch}{1.2}
 \tabcolsep=1.2pt
\caption{Typical ratios for $a_{1}(1930)$ and $a_1(2095)$. The $R$ ranges are $(4.58\sim4.92)$ GeV$^{-1}$ and $(4.78\sim5.16)$ GeV$^{-1}$ for $a_{1}(1930)$ and $a_1(2095)$, respectively.
 \label{a119302095r}}
\begin{tabular}{lcccccccc}
\toprule[1pt]
Ratio&$a_{1}(1930)$&$a_{1}(2095)$\\
\midrule[1pt]
$\Gamma_{\pi \rho}/\Gamma_{Total}$&$0.151\sim0.162$&$0.139\sim0.176$\\
$\Gamma_{\pi b_{1}(1235)}/\Gamma_{Total}$&$0.092\sim0.160$&$0.206\sim0.2542$\\
$\Gamma_{\pi \rho(1700)}/\Gamma_{Total}$&$0.005\sim0.024$&$0.039\sim0.0529$\\
$\Gamma_{\sigma \pi}/\Gamma_{Total}$&$0.088\sim0.097$&$0.058\sim0.073$\\
$\Gamma_{\pi \rho(1450)}/\Gamma_{Total}$&$0.339\sim0.347$&$0.189\sim0.253$\\
$\Gamma_{\pi b_{1}(1235)}/\Gamma_{\pi \rho(1450)}$&$0.271\sim0.462$&$0.348\sim1.813$\\
$\Gamma_{\eta a_{1}(1260)}/\Gamma_{KK^{*}(892)}$&$0.629\sim0.719$&$1.141\sim1.742$\\
$\Gamma_{\rho \omega}\Gamma_{\pi f_{2}(1270)}$&$0.705\sim0.850$&$0.188\sim0.451$\\
$\Gamma_{\eta a_{0}(980)}/\Gamma_{\pi \rho(1700)}$&$0.317\sim0.809$&$0.239\sim0.279$\\
$\Gamma_{KK_{1}(1400)}/\Gamma_{\eta a_{2}(1320)}$&$0.508\sim0.553$&$1.693\sim4.846$\\
$\Gamma_{KK_{1}(1400)}/\Gamma_{\rho a_{0}(980)}$&$-$&$0.145\sim0.184$\\
$\Gamma_{\eta a_{0}(1450)}/\Gamma_{KK_{0}^{*}(1430)}$&$-$&$0.206\sim0.838$\\
\bottomrule[1pt]
\end{tabular}
\end{table}

\begin{figure*}[htbp]
\includegraphics[scale=0.45]{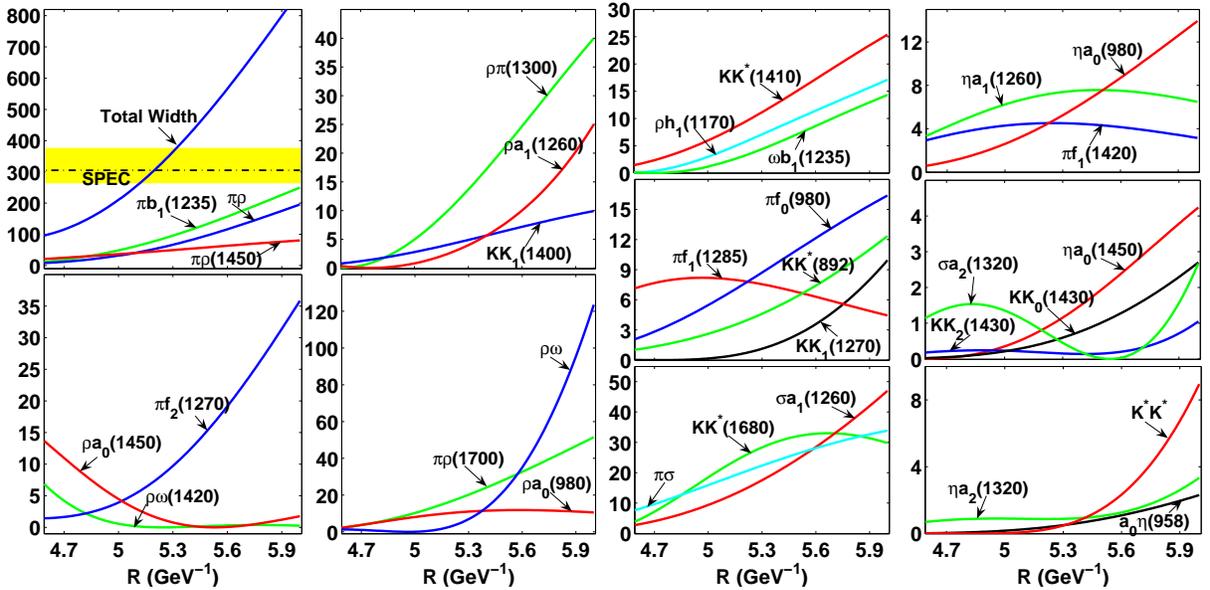}
\caption{(Color online) $R$ dependence of the calculated partial and total decay widths of $a_{1}(2270)$. Here, the dot-dashed line with the band is the experimental total width from Ref. \cite{Anisovich:2001pn}. All results are in units of MeV.
\label{a12270}}
\end{figure*}

In Fig. \ref{a12270}, we discuss the decay behavior of $a_{1}(2270)$
as the third radial excitation of $a_1(1260)$. 
{We find that the main decay mode includes the decay channels $\pi b_{1}(1235)$, $\pi\rho$, $\pi\rho(1450)$, and $\pi\rho(1700)$. In addition, $KK^{*}(1410)$, $\rho h_{1}(1170)$, $KK^{*}(1680)$, $\pi\sigma$, and $\sigma a_{1}(1260)$ have important contributions to the total decay width. $\rho a_{2}(1320)$, $\eta'(958)a_{0}(980)$, and $K^{*}K_{1}(1270)$ are subordinate decay modes}, which are not shown in Fig. \ref{a12270}.
In Table \ref{a12270r}, we also list the typical ratios of the decays of the $a_{1}(2270)$.

\begin{table}[htbp]
\renewcommand{\arraystretch}{1.2}
\tabcolsep=0.7pt
\caption{Calculated ratios of the decays of $a_{1}(2270)$. Here, all the results correspond to the $R$ range $(5.12\sim5.32)$ GeV$^{-1}$.
 \label{a12270r}}
\begin{tabular}{lcccccccc}
\toprule[1pt]
Ratio&Value&Ratio&Value\\
\midrule[1pt]
$\Gamma_{\pi \rho}/\Gamma_{Total}$&$0.164\sim0.184$&$\Gamma_{\pi f_{1}(1285)}/\Gamma_{\pi \sigma}$&$0.313\sim0.435$\\
$\Gamma_{\pi b_{1}(1235)}/\Gamma_{Total}$&$0.247\sim0.264$&$\Gamma_{KK^{*}(892)}/\Gamma_{\eta a_{1}(1260)}$&$0.313\sim0.487$\\
$\Gamma_{\pi \rho_{1700}}/\Gamma_{Total}$&$0.052\sim0.056$&$\Gamma_{\pi f_{1}(1420)}/\Gamma_{\rho a_{0}(980)}$&$0.404\sim0.469$\\
$\Gamma_{\sigma \pi}/\Gamma_{Total}$&$0.064\sim0.070$&$\Gamma_{\eta a_{0}(980)}/\Gamma_{\pi f_{2}(1270)}$&$0.532\sim0.612$\\
$\Gamma_{\pi \rho(1450)}/\Gamma_{Total}$&$0.134\sim0.157$&$\Gamma_{\eta a_{2}(1320)}/\Gamma_{\pi f_{0}}$&$0.099\sim0.131$\\
$\Gamma_{\rho a_{1}(1260)}/\Gamma_{\omega b_{1}(1235)}$&$0.789\sim0.926$&$\Gamma_{\eta a_{0}(1450)}/\Gamma_{\omega b_{1}(1235)}$&$0.236\sim0.273$\\
$\Gamma_{KK_{1}(1400)}/\Gamma_{\rho \pi_{1300}}$&$0.352\sim0.446$&$\Gamma_{\eta a_{1}(1260)}/\Gamma_{KK^{*}(1680)}$&$0.256\sim0.297$\\
$\Gamma_{\rho h_{1}(1170)}/\Gamma_{KK_{1}(1400)}$&$0.573\sim0.639$&$\Gamma_{KK^{*}(1410)}/\Gamma_{\sigma a_{1}(1260)}$&$0.633\sim0.638$\\
\bottomrule[1pt]
\end{tabular}
\end{table}

\subsection{$b_{1}$ states}

The Regge trajectory analysis indicates that $b_1(1235)$, $b_1(1960)$, and $b_1(2240)$ are the ground state,
second radial excitation, and {third} radial excitation in the $b_1$ meson family, respectively. In addition, we also predict a missing $b_1(1640)$ as the first radial excitation. In the following, we study their decays.

\begin{figure}[htbp]
\includegraphics[scale=0.55]{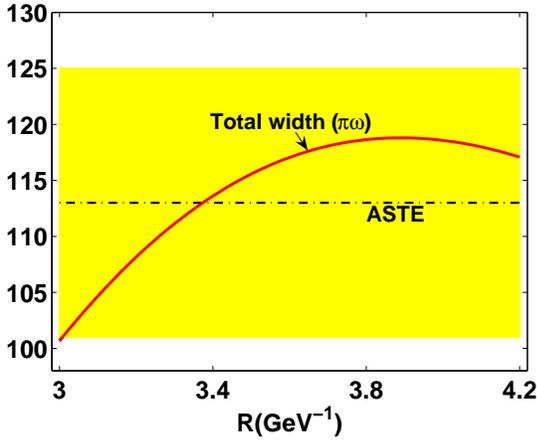}
\caption{(Color online) $R$ dependence of the calculated total decay width of $b_{1}(1235)$. Here, the dot-dashed line with the band is the experimental total width from Ref. \cite{Weidenauer:1993mv}. The total decay width is in units of
MeV. \label{b11235}}
\end{figure}

As for $b_{1}(1235)$, there are two allowed decay channels: $\pi\omega$ and $\pi a_0(980)$. The result shown in Fig. \ref{b11235} shows that the obtained
total width overlaps with experimental data from Ref. \cite{Weidenauer:1993mv}. Since $b_{1}\rightarrow\omega\pi$ occurs via $S$ and $D$ waves, we obtain
the $D$-wave/$S$-wave amplitude ratio of the $b_{1}\rightarrow\omega\pi$ process, which is $0.465$ in our work; this is consistent with the Crystal Barrel data
($0.45\pm0.04$) \cite{Amsler:1993pr}. On the other hand, the decay channel $\pi f_{0}$ has a partial decay width that is less than 1 MeV.

\begin{figure}[htbp]
\includegraphics[scale=0.45]{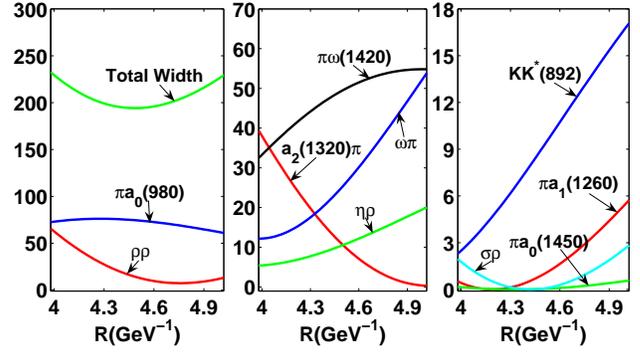}
\caption{(Color online) $R$ dependence of the calculated partial and total decay widths of $b_{1}(1640)$. All results are in units of MeV. \label{b11640}}
\end{figure}

As a predicted $b_1$ state, $b_{1}(1640)$ has the decay behavior listed in Fig. \ref{b11640}, where we take the same $R$ range as that for $a_{1}(1640)$. \footnote{Since $b_1(1640)$ is a predicted state, we take the same $R$ range as that of  $a_1(1640)$ to predict the decay behavior of $b_1(1640)$. This treatment is due to the fact that $b_1(1640)$ is the isospin partner of  $a_1(1640)$, which has a similar $R$ range.} Its main decay channel is $\pi a_{0}(980)$, while $\pi a_{2}(1320)$, $\rho \rho$, $\pi\omega(1420)$, $KK^{*}$, and $\omega \pi$ also have considerable
contributions to the total decay width. The total decay width is predicted to be $200\sim232$ MeV. Table \ref{b11640r} shows some ratios that are relevant to the decays of $b_{1}(1640)$, which is valuable for further experimental searches for this axial-vector state.

\renewcommand{\arraystretch}{1.2}
\begin{table}[htbp]
\caption{Typical ratios for decays of $b_1(1640)$ corresponding to $R=4.20\sim4.90$ GeV$^{-1}$.
 \label{b11640r}}
\begin{tabular}{lccc}
\toprule[1pt]
Ratio&Value&Ratio&Value\\
\midrule[1pt]
$\Gamma_{\pi a_{0}(980)}/\Gamma_{Total}$&$0.352\sim0.368$&$\Gamma_{KK^{*}}/\Gamma_{\omega\pi}$&$0.324\sim0.347$\\
$\Gamma_{\eta\rho}/\Gamma_{\pi\omega(1420)}$&$0.164\sim0.263$&$\Gamma_{\pi a_{2}(1320)}/\Gamma_{\rho\rho}$&$0.565\sim0.681$\\
\bottomrule[1pt]
\end{tabular}
\end{table}

\begin{figure}[htbp]
\includegraphics[scale=0.39]{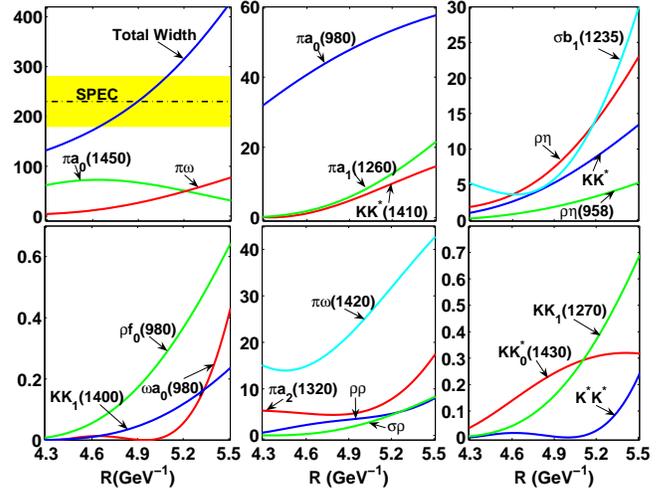}
\caption{(Color online) $R$ dependence of the calculated partial and total decay widths of $b_{1}(1960)$. Here, the dot-dashed line with the band is the
experimental total width from Ref. \cite{Anisovich:2002su}. Since the width of the $KK^{*}_{2}$ mode is tiny, we do not list its contribution here.
All results are in units of MeV. \label{b11960}}
\end{figure}

Assuming that $b_{1}(1960)$ is the second radial excitation of $b_1(1235)$, we present its total and partial decay widths in Fig. \ref{b11960}. Our calculated
total width can cover the experimental data given in Ref. \cite{Anisovich:2002su}. Its main decay channels are $\pi a_{0}(1450)$, $\pi\omega$, $\pi
a_{0}(980)$, and $\pi\omega(1420)$, while the partial decay widths of the decay modes $\pi a_{1}(1260)$, $\rho\eta$, and $\pi a_{2}(1320)$ are also
considerable. We also obtain some ratios of partial decay widths of $b_{1}(1960)$,which are listed in Table \ref{b11960r}.

\begin{table}[htbp]

\caption{Obtained ratios for decays of $b_{1}(1960)$. All results correspond to $R=4.66\sim5.16$ GeV$^{-1}$.
 \label{b11960r}}
\begin{tabular}{lccc}
\toprule[1pt]
Ratio&Value&Ratio&Value\\
\midrule[1pt]
$\Gamma_{\pi a_{0}(980)}/\Gamma_{Total}$&$0.186\sim0.235$&$\Gamma_{\rho \rho}/\Gamma_{Total}$&$0.028\sim0.031$\\
$\Gamma_{\pi \omega_{1420}}/\Gamma_{Total}$&$0.088\sim0.107$&$\Gamma_{\omega \pi}/\Gamma_{Total}$&$0.077\sim0.162$\\
$\Gamma_{\rho \rho}/\Gamma_{\pi a_{2}(1320)}$&$0.572\sim0.624$&$\Gamma_{KK^{*}(892)}/\Gamma_{\eta \rho}$&$0.648\sim0.736$\\
$\Gamma_{\rho f_{0}}/\Gamma_{\pi a_{1}(1260)}$&$0.029\sim0.030$&$\Gamma_{KK_{1}(1400)}/\Gamma_{KK_{1}(1270)}$&$0.249\sim0.316$\\
\bottomrule[1pt]
\end{tabular}
\end{table}

\begin{figure*}[htbp]
\includegraphics[scale=0.45]{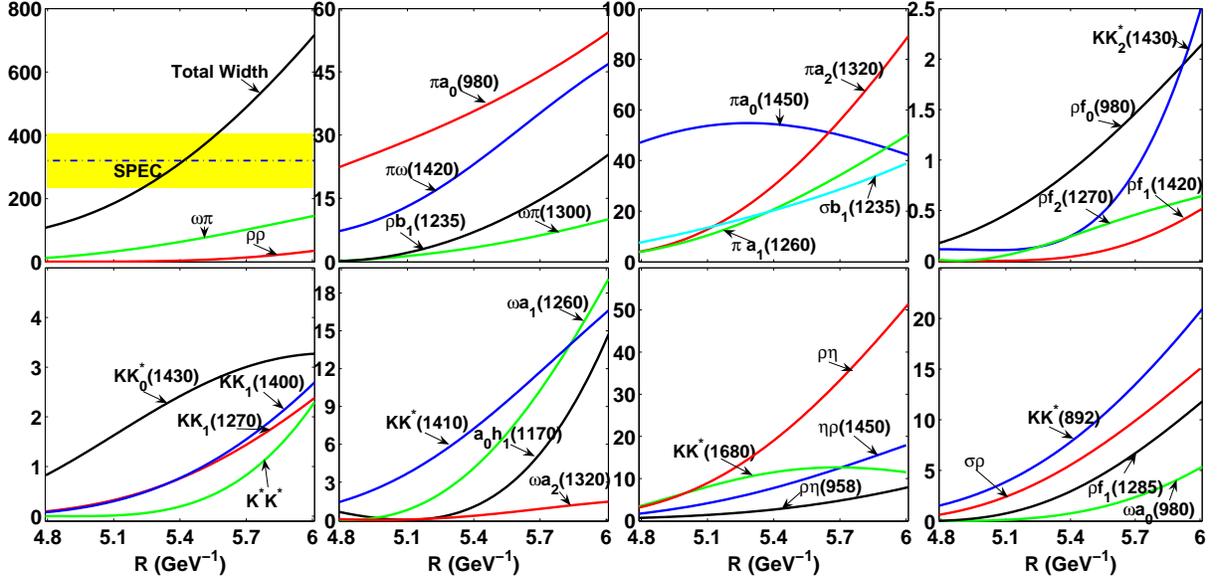}
\caption{(Color online) $R$ dependence of the calculated partial and total decay widths of $b_{1}(2240)$. Here, the dot-dashed line with the band
is the experimental total width from Ref. \cite{Anisovich:2002su}. We do not present the $K^{*} K_{1}(1270)$ contribution since this decay has a tiny width.
All results are in units of MeV. \label{b12240}}
\end{figure*}

In Fig. \ref{b12240}, we show the decay behavior of $b_{1}(2240)$ as the third radial excitation of $b_1(1235)$.
Additionally, its main decay modes are $\omega\pi$, $\pi\omega(1420)$, $\pi a_{0}(980)$, and $\pi a_{0}(1450)$. Of course, the decay modes $\rho\rho$, $\rho b_{1}(1235)$, $\pi
a_{2}(1320)$, and $\pi a_{1}(1260)$ also have obvious contributions to the total decay width. For the convenience of further experimental studies of this
state, we provide information on typical ratios of
the partial width of $b_{1}(2240)$ in Table \ref{b12240r}.

\begin{table}[htbp]

 \tabcolsep=0.7pt
\caption{Calculated ratios for $b_{1}(2240)$ corresponding to $R=5.20\sim5.54$ GeV$^{-1}$.
 \label{b12240r}}
\begin{tabular}{lccc}
\toprule[1pt]
Ratio&Value&Ratio&Value\\
\midrule[1pt]
$\Gamma_{\pi a_{0}(980)}/\Gamma_{Total}$&$0.097\sim0.128$&$\Gamma_{\pi a_{0}(1450)}/\Gamma_{Total}$&$0.131\sim0.232$\\
$\Gamma_{\pi \omega_{1420}}/\Gamma_{Total}$&$0.068\sim0.071$&$\Gamma_{\omega \pi}/\Gamma_{Total}$&$0.179\sim0.199$\\
$\Gamma_{\pi a_{2}(1320)}/\Gamma_{Total}$&$0.075\sim0.102$&$\Gamma_{\pi a_{1}(1260)}$&$0.057\sim0.066$\\
$\Gamma_{\eta \rho(1450)}/\Gamma_{Total}$&$0.055\sim0.064$&$\Gamma_{\eta \rho}/\Gamma_{Total}$&$0.050\sim0.062$\\
$\Gamma_{KK^{*}(1680)}/\Gamma_{Total}$&$0.042\sim0.057$&$\Gamma_{\rho f_{1}(1285)}/\Gamma_{\omega a_{1}(1260)}$&$0.678\sim0.819$\\
$\Gamma_{\rho f_{0}}/\Gamma_{\rho b_{1}(1235)}$&$0.112\sim0.173$&$\Gamma_{\rho f_{2}(1270)}/\Gamma_{KK_{1}(1400)}$&$0.254\sim0.317$\\
$\Gamma_{KK_{1}(1270)}/\Gamma_{KK^{*}_{0}(1430)}$&$0.233\sim0.383$&$\Gamma_{\rho \eta'(958)}/\Gamma_{KK^{*}(1410)}$&$0.364\sim0.381$\\
$\Gamma_{KK^{*}(1410)}/\Gamma_{KK^{*}(892)}$&$0.903\sim0.950$&$\Gamma_{\omega a_{0}(980)}/\Gamma_{\omega a_{1}(1260)}$&$0.158\sim0.204$\\
\bottomrule[1pt]
\end{tabular}
\end{table}

\subsection{$f_{1}$ states}

When discussing $f_1$ states, we need to consider the admixtures of the flavor wave functions $|n\bar{n}\rangle=(|u\bar{u}\rangle+|d\bar{d}\rangle)/\sqrt{2}$
and $|s\bar{s}\rangle$. $f_1(1285)$ and $f_1(1420)/f_1(1510)$ satisfy
%
\begin{equation}
 \left(
  \begin{array}{c}
   |f_{1}(1285)\rangle\\
   |f_1(1420)/f_1(1510)\rangle\\
  \end{array}
\right )=
\left(
  \begin{array}{cc}
    \cos\phi & -\sin\phi \\
   \sin\phi & \cos\phi\\
  \end{array}
\right)
\left(
  \begin{array}{c}
    |n\bar{n}  \rangle \\
    |s\bar{s} \rangle\\
  \end{array}
\right),\label{m1}
\end{equation}
where both $f_1(1420)$ and $f_1(1510)$ are partners of $f_1(1285)$. (We present their decay behaviors below.) $\phi$ denotes a mixing angle. This
mixing angle was determined in a phenomenological way \cite{Close:1997nm} and is given by $\phi=(20-30)^{\circ}$, which is consistent with $\phi=(24^{+3.2}_{-2.7})^{\circ}$
reported by the LHCb Collaboration \cite{Aaij:2013rja} and $\phi=(21\pm5)^{\circ}$ from the updated lattice QCD analysis \cite{Dudek:2013yja}. When
calculating the decays of $f_1(1285)$ and $f_1(1420)/f_1(1510)$, we take the LHCb value $\phi=24^\circ$.

In Fig. \ref{Regge}, we have predicted that $f_1(1640)$ is the first radial excitation of $f_1(1285)$ and that $f_1(1800)$ is a partner of $f_1(1640)$; these two predicted axial-vector mesons are related by
%
\begin{equation}
 \left(
  \begin{array}{c}
   |f_{1}(1640)\rangle\\
   |f_1(1800)\rangle\\
  \end{array}
\right )=
\left(
  \begin{array}{cc}
    \cos\phi_1 & -\sin\phi_1 \\
   \sin\phi_1 & \cos\phi_1\\
  \end{array}
\right)
\left(
  \begin{array}{c}
    |n\bar{n}  \rangle \\
    |s\bar{s} \rangle\\
  \end{array}
\right),\label{m1}
\end{equation}
In addition, there exist relations among $f_1(1970)$, the predicted $f_1(2110)$ and $f_1(2210)$, and $f_1(2310)$, i.e.,
%
\begin{equation}
 \left(
  \begin{array}{c}
   |f_{1}(1970)\rangle\\
   |f_1(2110)\rangle\\
  \end{array}
\right )=
\left(
  \begin{array}{cc}
    \cos\phi_2 & -\sin\phi_2 \\
    \sin\phi_2 & \cos\phi_2\\
  \end{array}
\right)
\left(
  \begin{array}{c}
    |n\bar{n}  \rangle \\
    |s\bar{s} \rangle\\
  \end{array}
\right),\label{m2}
\end{equation}
and
%
\begin{equation}
 \left(
  \begin{array}{c}
   |f_{1}(2210)\rangle\\
   |f_1(2310)\rangle\\
  \end{array}
\right )=
\left(
  \begin{array}{cc}
    \cos\phi_3 & -\sin\phi_3 \\
    \sin\phi_3 & \cos\phi_3\\
  \end{array}
\right)
\left(
  \begin{array}{c}
    |n\bar{n}  \rangle \\
    |s\bar{s} \rangle\\
  \end{array}
\right),\label{m3}
\end{equation}
Here, the mixing angles $\phi_i$ ($i=1,2,3$) cannot be constrained by our analysis. In the following discussions, we take a typical value $\phi_i=\phi=24^\circ$ to
give the quantitative results.

\begin{figure}[htbp]
\includegraphics[scale=0.48]{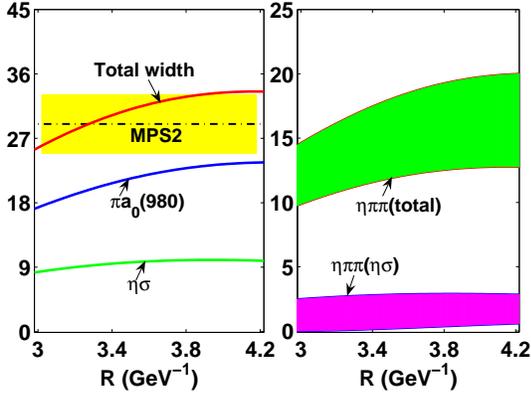}
\caption{(Color online) $R$ dependence of the total and partial decay widths of $f_{1}(1285)$.
We also present the decay width of $f_{1}(1285)\to \eta\pi\pi$ via the intermediate channels $\eta\sigma$ and $\pi a_{0}(980)$ (green band), and only from
the intermediate channel $\eta\sigma$ (pink band).
Here, the experimental total width from Ref. \cite{Lee:1994bh} is denoted by the dot-dashed line with the band. All results are in units of MeV. \label{f11285}}
\end{figure}

As for $f_{1}(1285)$, we show its partial and total decay widths in Fig. \ref{f11285}, where the calculated total decay width is in agreement with the
experimental data from Ref. \cite{Lee:1994bh}.
However, we notice that the calculated branching ratio for $\Gamma_{\pi a_{0}}/\Gamma_{total}=0.67\sim0.68$, corresponding to $R=(3.00\sim4.00)$ GeV$^{-1}$,
which is a little bit larger than $(36\pm7)\%$ listed in the PDG \cite{Beringer:1900zz}. The PDG data also shows that the branching ratio of its decay $\eta\pi\pi$
can reach up to $(52.4^{+1.9}_{-2.2})\%$ \cite{Beringer:1900zz}, which is the main contribution to the total decay width of $f_{1}(1285)$. {In this
work, we study the processes $f_{1}(1285)\rightarrow\eta\sigma\rightarrow\eta\pi\pi$ and $f_{1}(1285)\rightarrow\pi a_{0}(980)\rightarrow\eta\pi\pi$, which
can be calculated using the QPC model. Thus, the decay width of $f_{1}(1285)\rightarrow\eta\sigma\rightarrow\eta\pi\pi$ can be written as \cite{Luo:2009wu}
\begin{eqnarray}
&&\Gamma(f_{1}\to \eta+\sigma\to\eta+\pi\pi)\nonumber\\&&=\frac{1}{\pi}
\int_{4m_\pi^2}^{(m_{f_{1}}-m_\eta)^2}dr\sqrt{r}\frac{\Gamma_{f_{1}\to \eta+\sigma}(r)\cdot \Gamma_{\sigma\to
\pi\pi}(r)}{(r-{m_{\sigma}^2})^2+(m_{\sigma}\Gamma_{\sigma})^2}\label{11},
\end{eqnarray}
where the interaction of $\sigma$ with two pions can be described by the effective Lagrangian
\begin{equation}
\mathcal{L}_{\sigma\pi\pi}=g_{\sigma}\sigma(2\pi^{+}\pi^{-}+\pi^{0}\pi^{0}).
\end{equation}
The coupling constant $g_{\sigma}=2.12\sim 2.81$ GeV is determined by the total width $\Gamma_{\sigma}=400\sim 700$ MeV \cite{Beringer:1900zz}, and the decay width reads as
\begin{equation}
\Gamma_{\sigma\to\pi\pi}(r)=\frac{g_{\sigma}^{2}\lambda^{2}}{8\pi r}\frac{[(r-(2m_{\pi})^{2})r]^{1/2}}{2\sqrt{r}},
\end{equation}
where $\lambda=\sqrt{2}$ and $1$ for $\pi^{+}\pi^{-}$ and $\pi^{0}\pi^{0}$, respectively.

The process $f_{1}(1285)\rightarrow\pi a_{0}(980)\rightarrow\eta\pi\pi$ is calculated in a similar way, and the equation is given by
\begin{eqnarray}
&&\Gamma(f_{1}\to a_{0}+\pi\to\eta+\pi\pi)\nonumber\\&&=\frac{1}{\pi}
\int_{(m_\pi+m_\eta)^2}^{(m_{f_{1}}-m_\pi)^2}dr\sqrt{r}\frac{\Gamma_{f_{1}\to \pi+a_0}(r)\cdot \Gamma_{a_0\to
\eta\pi}(r)}{(r-{m_{a_0}^2})^2+(m_{a_0}\Gamma_{a_0})^2}\label{11},
\end{eqnarray}
where the decay width for $a_{0}(980)\rightarrow\eta\pi$ is
\begin{eqnarray}
&&\Gamma_{a_0(980)\to\eta\pi}(r)\nonumber\\&&=\frac{g_{a_0}^{2}}{8\pi r}\frac{[(r-(m_{\eta}+m_{\pi})^{2})(r-(m_{\eta}-m_{\pi}))^{2}]^{1/2}}{2\sqrt{r}},
\end{eqnarray}
where the coupling constant $g_{a_0}=1.262\sim2.524$ GeV is determined by the total width of $a_0(980)$ ($\Gamma_{a_0(980)}=50\sim100$ MeV). The final
result of the width of ${f_{1}(1285)\rightarrow\pi a_{0}(980)\rightarrow\eta\pi\pi}$ includes the contributions from both $\eta\pi^{0}\pi^{0}$ and
$\eta\pi^{+}\pi^{-}$.

The decay width of $f_{1}(1285)\rightarrow\eta\pi\pi$ via both the intermediate $\eta\sigma$ and $\pi a_{0}(980)$ channels and only the intermediate
$\eta\sigma$ channel are shown in Fig. \ref{f11285}. In addition, the decay width of $f_{1}(1285)\rightarrow\eta\pi\pi$ from the intermediate $\pi
a_{0}(980)$ channel is comparable with the corresponding experimental data $[(16\pm7)\%]$ in the PDG \cite{Beringer:1900zz}.}

\begin{figure}[htbp]
\includegraphics[scale=0.48]{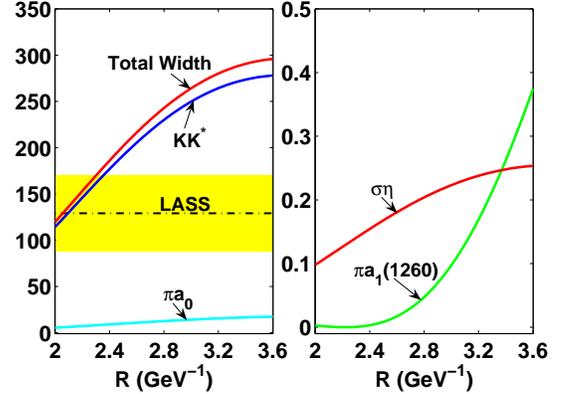}
\caption{(Color online) $R$ dependence of the total and partial decay widths of $f_{1}(1420)$. Here, the experimental total width from Ref. \cite{Augustin:1990ki} is shown by the
dot-dashed line with the band. All results are in units of MeV. \label{f1420}}
\end{figure}

\begin{figure}[htbp]
\includegraphics[scale=0.4]{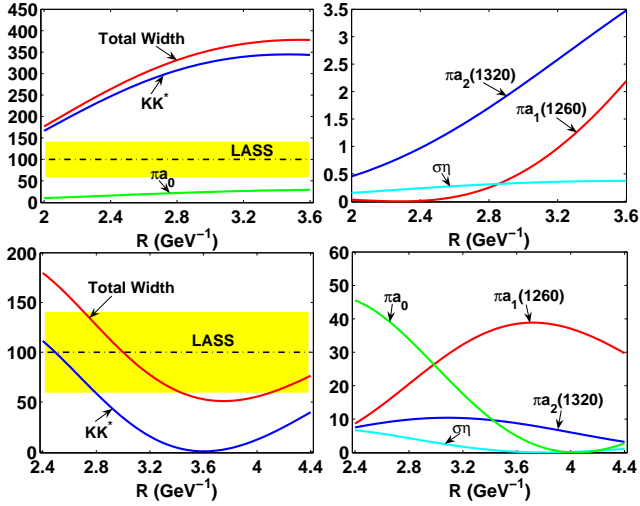}
\caption{(Color online) Total and partial decay widths of $f_{1}(1510)$ as a partner of $f_1(1285)$ (first row) and as the first radial
excitation of $f_1(1285)$ (second row). The experimental total width from Ref. \cite{Aihara:1988bw} is denoted by the
dot-dashed line with the band. All results are in units of MeV. \label{f11510n1}}
\end{figure}

In the following, we discuss the decay behaviors of $f_1(1420)$ and $f_1(1510)$ as partners of $f_1(1285)$. As for $f_1(1420)$, the obtained total decay
width can overlap with the DM2 result \cite{Augustin:1990ki}, as shown in Fig. \ref{f1420}. Its main decay channel is $KK^{*}$. Thus, the present study of
decay of $f_1(1420)$ supports the prediction that $f_1(1420)$ is a partner of $f_1(1285)$. As for $f_1(1510)$, its partial and total decay widths are listed in Fig
.\ref{f11510n1}, which shows that the calculated total decay width is larger than the experimental data \cite{Augustin:1990ki}. Thus, the possibility that $f_1(1510)$ is a partner
of $f_1(1285)$ can be excluded.


\begin{figure}[htbp]
\includegraphics[scale=0.40]{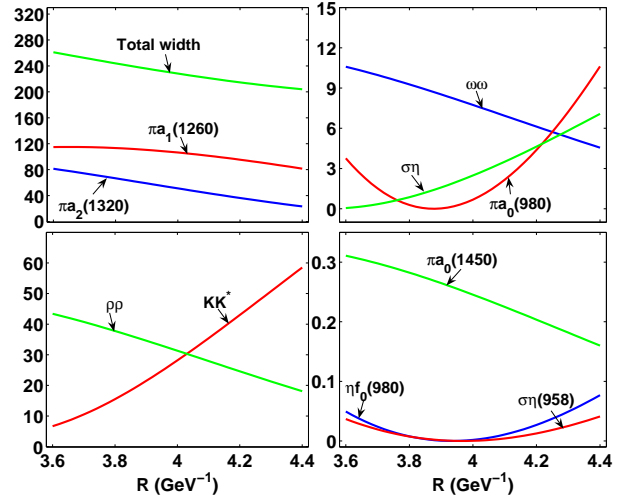}
\caption{(Color online) $R$ dependence of the total and partial decay widths of $f_{1}(1640)$. All results are in units of MeV.
\label{f11640}}
\end{figure}

\begin{figure}[htbp]
\includegraphics[scale=0.40]{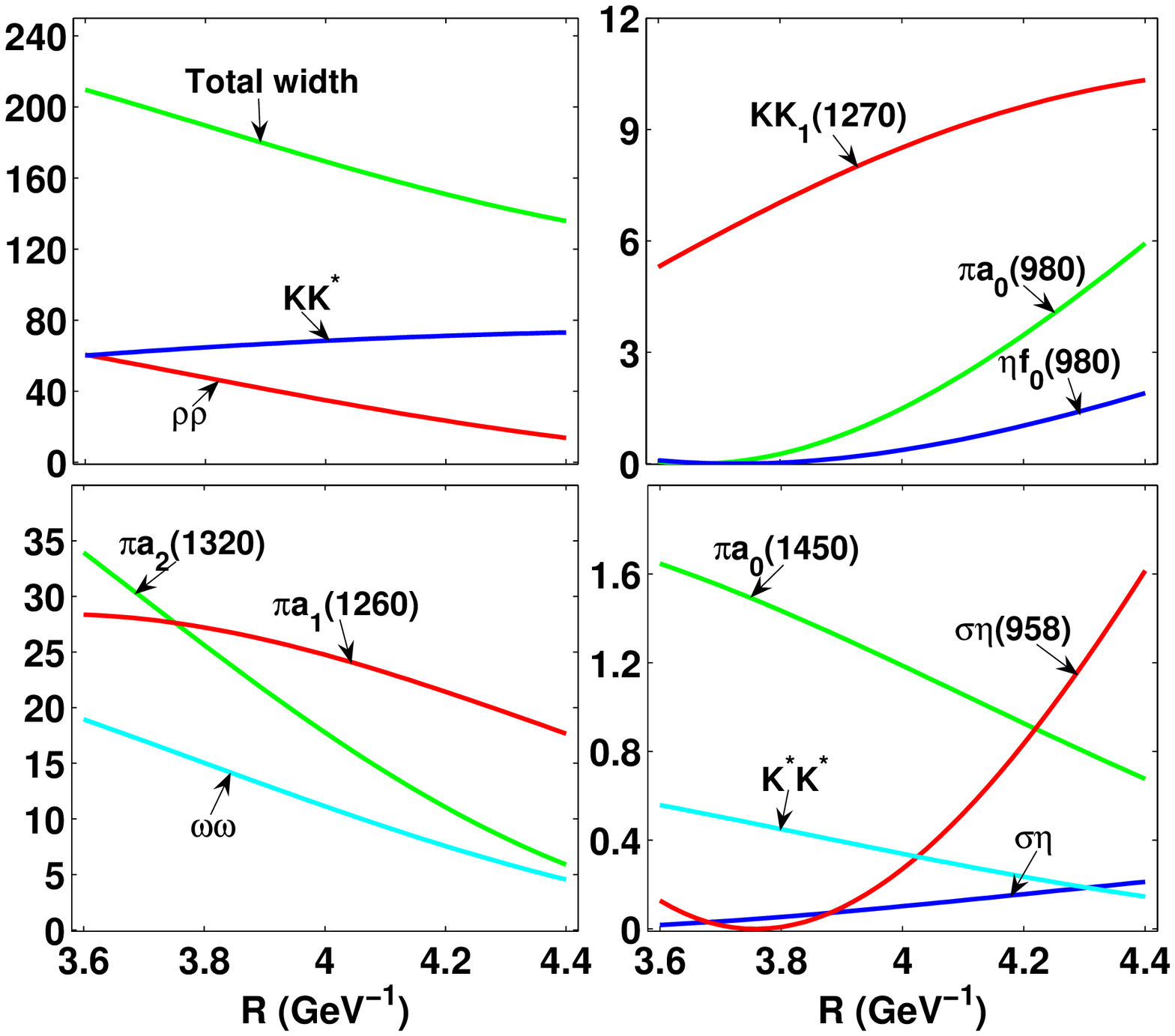}
\caption{(Color online) $R$ dependence of the total and partial decay widths of $f_{1}(1800)$. All results are in units of MeV.
\label{f11800}}
\end{figure}

In Figs. \ref{f11640} and \ref{f11800}, we further illustrate the decay properties of the two predicted states $f_{1}(1640)$ and $f_{1}(1800)$.
In addition, we also list some of their typical ratios, which are weakly dependent on the $R$ value (see Table \ref{f2p branching}), where we take
$R=(3.60\sim4.40)$ GeV$^{-1}$. From Figs. \ref{f11640} and \ref{f11800} and Table \ref{f2p branching}, we can obtain information on the main decay modes and
the resonance parameters of the two predicted $f_1$ mesons.

As for $f_{1}(1510)$, there also exists another possible assignment, i.e., $f_{1}(1510)$ can be a radial excitation of $f_1(1285)$ since the
mass of $f_{1}(1510)$ is close to that of the predicted $f_1(1640)$. Here, we use the mixing angle expression
\begin{equation}
|f_{1}(1510)\rangle=\cos\phi_{1}|n\bar{n}\rangle-\sin\phi_{1}|s\bar{s}\rangle,
\end{equation}
which is the same as $f_{1}(1640)$. Thus, we also further illustrate the decay behavior of $f_{1}(1510)$ as a radial
excitation of $f_1(1285)$ (see Fig. \ref{f11510n1}). Under this assignment, the obtained total decay width can be fitted with the LASS data \cite{Aihara:1988bw}. The $KK^{*}$ mode also has a large contribution to the total decay width. These facts indicate the possibility that $f_{1}(1510)$ is a radial excitation of $f_1(1285)$.

\begin{table}[tbp]
\caption{{Some obtained ratios relevant to decays of $f_{1}(1640)$ and $f_{1}(1800)$. All values correspond to the $R$ range $(3.60\sim4.40)$
GeV$^{-1}$.}
 \label{f2p branching}}
\begin{tabular}{cc|cc}

\toprule[1pt]
\multicolumn{2}{c|}{$f_{1}(1640)$} &\multicolumn{2}{c}{$f_{1}(1800)$}\\
\midrule[1pt]
$\Gamma_{\pi a_{1}(1260)}/\Gamma_{Total}$&$0.400\sim0.440$ &$\Gamma_{\rho\rho}/\Gamma_{Total}$&$0.102\sim0.290$\\
$\Gamma_{\pi a_{2}(1320)}/\Gamma_{Total}$&$0.114\sim0.312$ &$\Gamma_{\omega\omega}/\Gamma_{\pi a_{1}(1260)}$&$0.254\sim0.665$\\
$\Gamma_{\omega\omega}/\Gamma_{\rho\rho}$&$0.244\sim0.254$&$\Gamma_{KK_{1}(1270)}/\Gamma_{KK^{*}}$&$0.088\sim0.141$\\
$\Gamma_{KK^{*}}/\Gamma_{Total}$&$0.026\sim0.284$&$\Gamma_{\pi a_{2}(1320)}/\Gamma_{Total}$&$0.043\sim0.162$\\
\bottomrule[1pt]
\end{tabular}
\end{table}

\begin{figure}[htbp]
\includegraphics[scale=0.4]{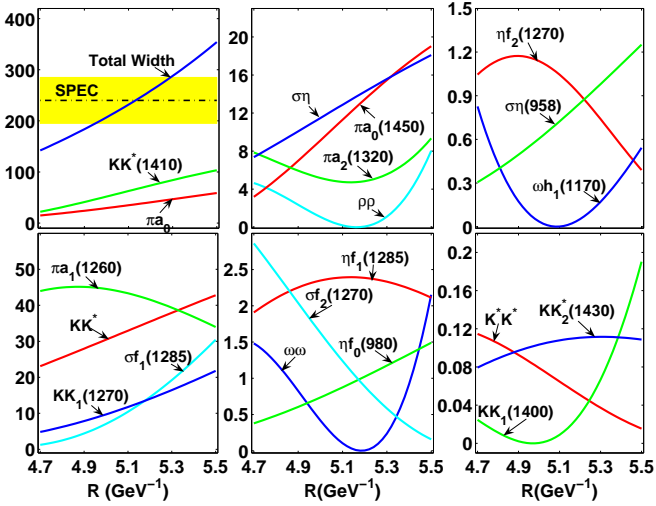}
\caption{(Color online) $R$ dependence of the total and partial decay widths of $f_{1}(1970)$. The experimental total width from Ref.
\cite{Anisovich:2000ut} is denoted by the dot-dashed line with the band. All results are in units of MeV. \label{f11970}}
\end{figure}

\begin{figure*}[htbp]
\includegraphics[scale=0.45]{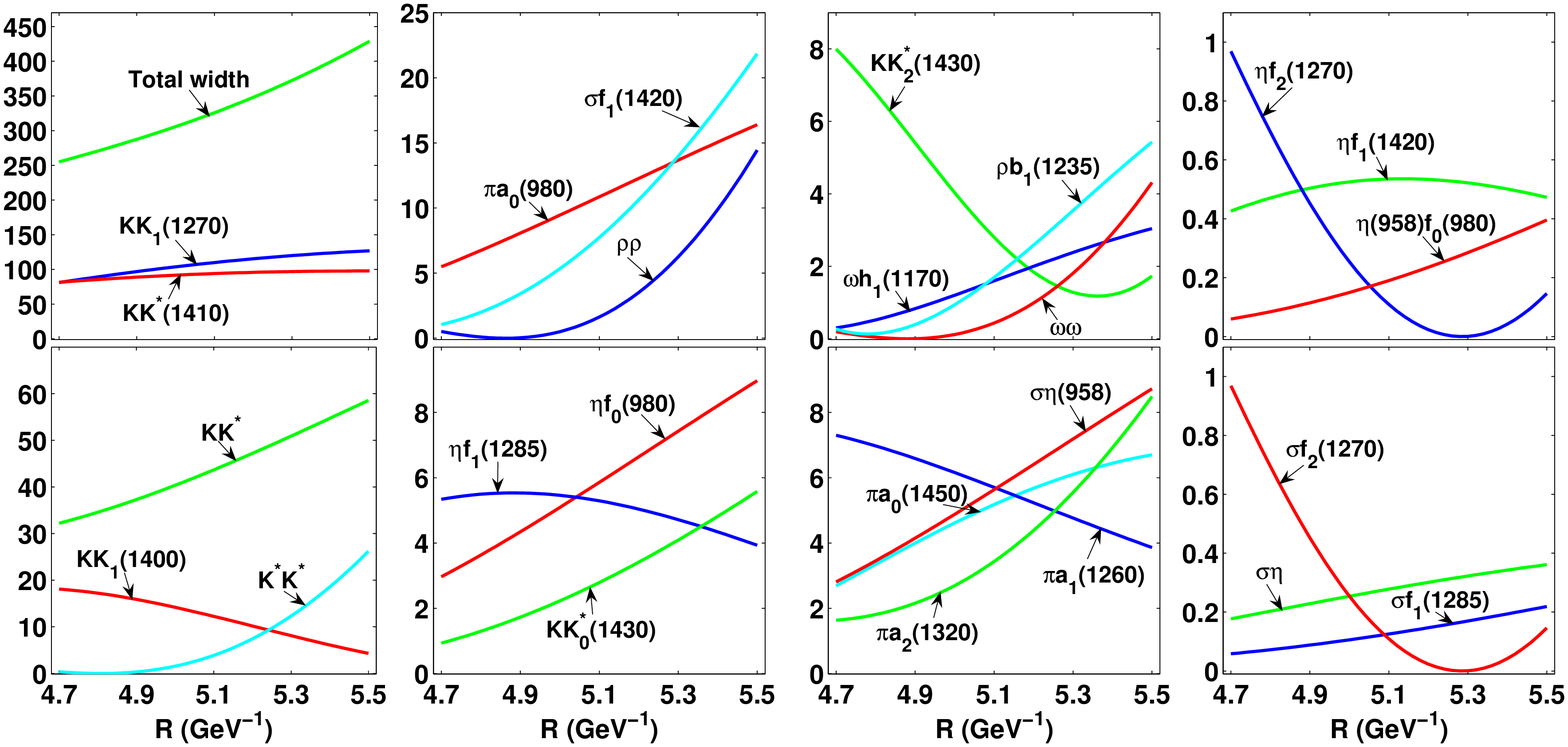}
\caption{(Color online) $R$ dependence of the total and partial decay widths of $f_{1}(2110)$. All results are in units of MeV. \label{f12110}}
\end{figure*}

In Fig. \ref{f11970}, we show the $R$ dependence of the decay behavior of $f_{1}(1970)$ as the second radial excitation of $f_1(1285)$. 
Its main decay channels are $KK^{*}(1410)$, $\pi a_{0}(980)$, $\pi a_{1}(1260)$, and $KK^{*}$. As a partner of $f_{1}(1970)$, the predicted
$f_{1}(2110)$ mainly decays into $KK_{1}(1270)$, $KK^{*}(1410)$, and $KK^{*}$ and has a large total decay width, as shown in Fig. \ref{f12110}.

\begin{figure*}[htbp]
\includegraphics[scale=0.45]{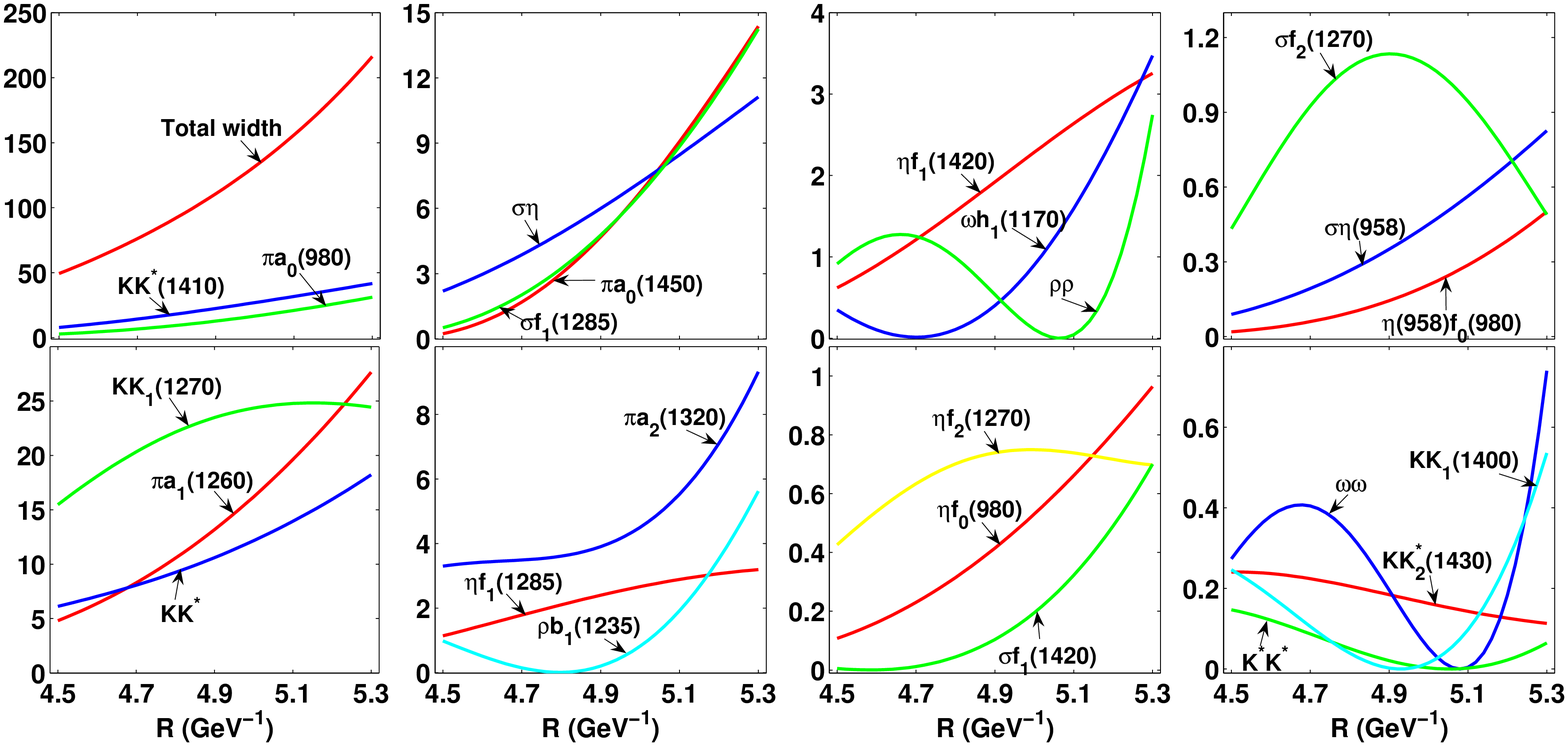}
\caption{(Color online) $R$ dependence of the total and partial decay widths of $f_{1}(2210)$. All results are in units of MeV.
\label{f12210}}
\end{figure*}

\begin{figure*}[htbp]
\includegraphics[scale=0.45]{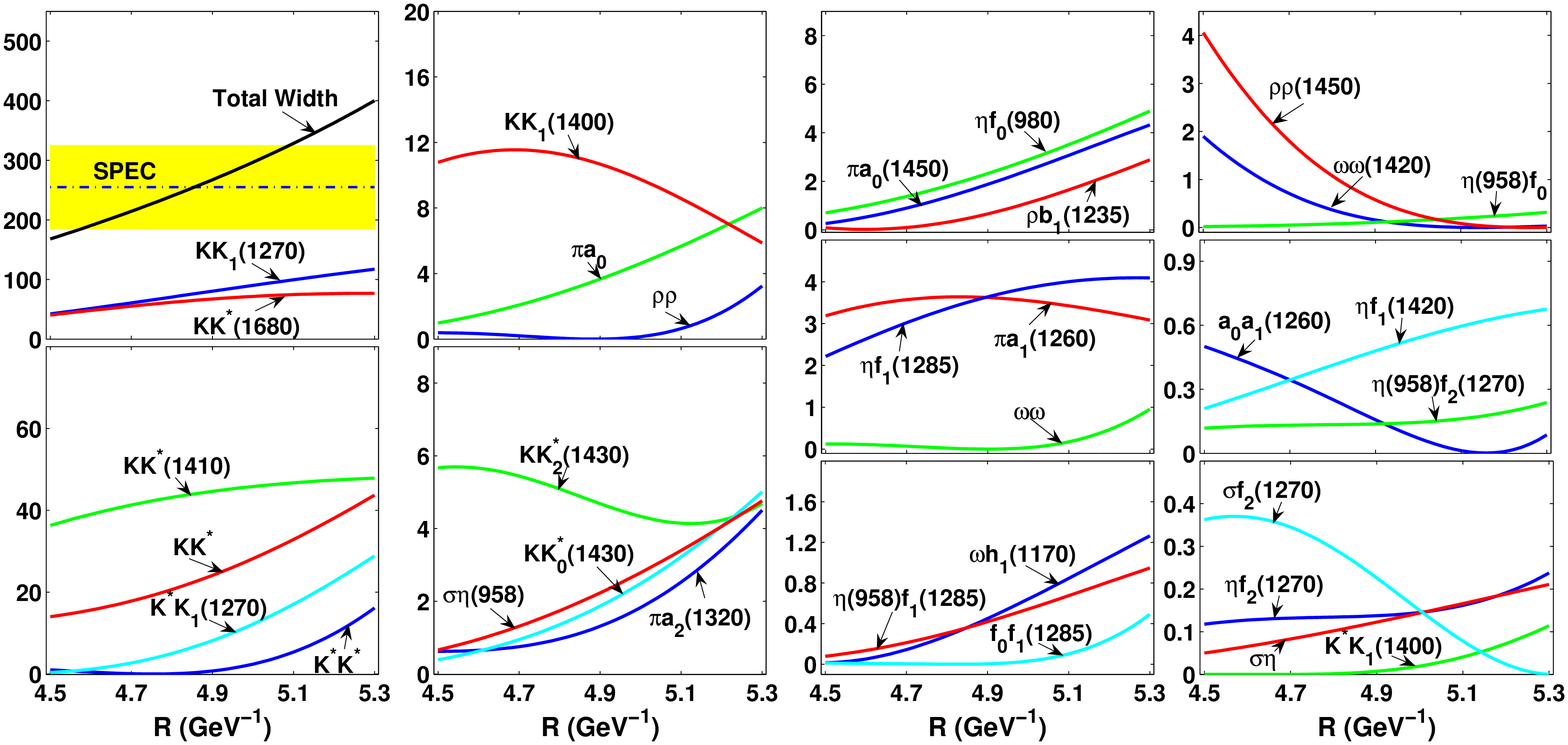}
\caption{(Color online) $R$ dependence of the total and partial decay widths of $f_{1}(2310)$. The experimental total width from Ref.
\cite{Anisovich:2000ut} is denoted by the dot-dashed line with the band. All results are in units of MeV. \label{f12310}}
\end{figure*}

The third radial excitation of $f_1(1285)$ is still missing in experiments. In this work, we predict $f_{1}(2210)$, and we calculate its total and partial decay widths (see Fig. \ref{f12210}). As a partner of this predicted $f_1(2210)$, $f_{1}(2310)$ has the decay properties listed in
Fig. \ref{f12310}, in which the experimental width \cite{Anisovich:2000ut} is depicted by our calculation when taking $R=(4.58\sim5.10)$
GeV$^{-1}$. Its main decay channels are $KK_{1}(1270)$, $KK^{*}(1680)$, $KK^{*}(1410)$, and $KK^{*}$.

\subsection{$h_{1}$ states}

Similar to the $f_1$ mesons, the following study of $h_1$ states is relevant to the admixtures of the flavor wave functions $n\bar{n}$ and $s\bar{s}$.
As the ground states in the $h_1$ meson family, $h_{1}(1170)$ and $h_{1}(1380)$ satisfy
%
{
\begin{eqnarray}
&&\left(
  \begin{array}{c}
   |h_1(1170)\rangle\\
   -|h_1(1380)\rangle\\
  \end{array}
\right )=
\left(
  \begin{array}{cc}
    \sin\theta_1 & \cos\theta_1 \\
    -\cos\theta_1 & \sin\theta_1\\
  \end{array}
\right)
\left(
  \begin{array}{c}
    |n\bar{n}  \rangle \\
    |s\bar{s} \rangle\\
  \end{array}
\right),
\nonumber \\\label{h1eq1}
\end{eqnarray}
%
%
}where the mixing angle $\theta_1$ is introduced, the first line of this equation is adopted in this paper, and the second line is used in Ref. \cite{Cheng:2011pb}. {In Ref. \cite{Cheng:2011pb},  Cheng obtained $\theta_1\sim82.7^{\circ}$. The lattice QCD calculation
indicates $\theta_1=86.8^{\circ}$ \cite{Dudek:2011tt}. In addition, $\theta_1=85.6^{\circ}$ was obtained in Ref. \cite{Li:2005eq}. In our
calculation, we present our result as {$\theta_1=85.6^\circ$}.}

\begin{figure}[htbp]
\includegraphics[scale=0.45]{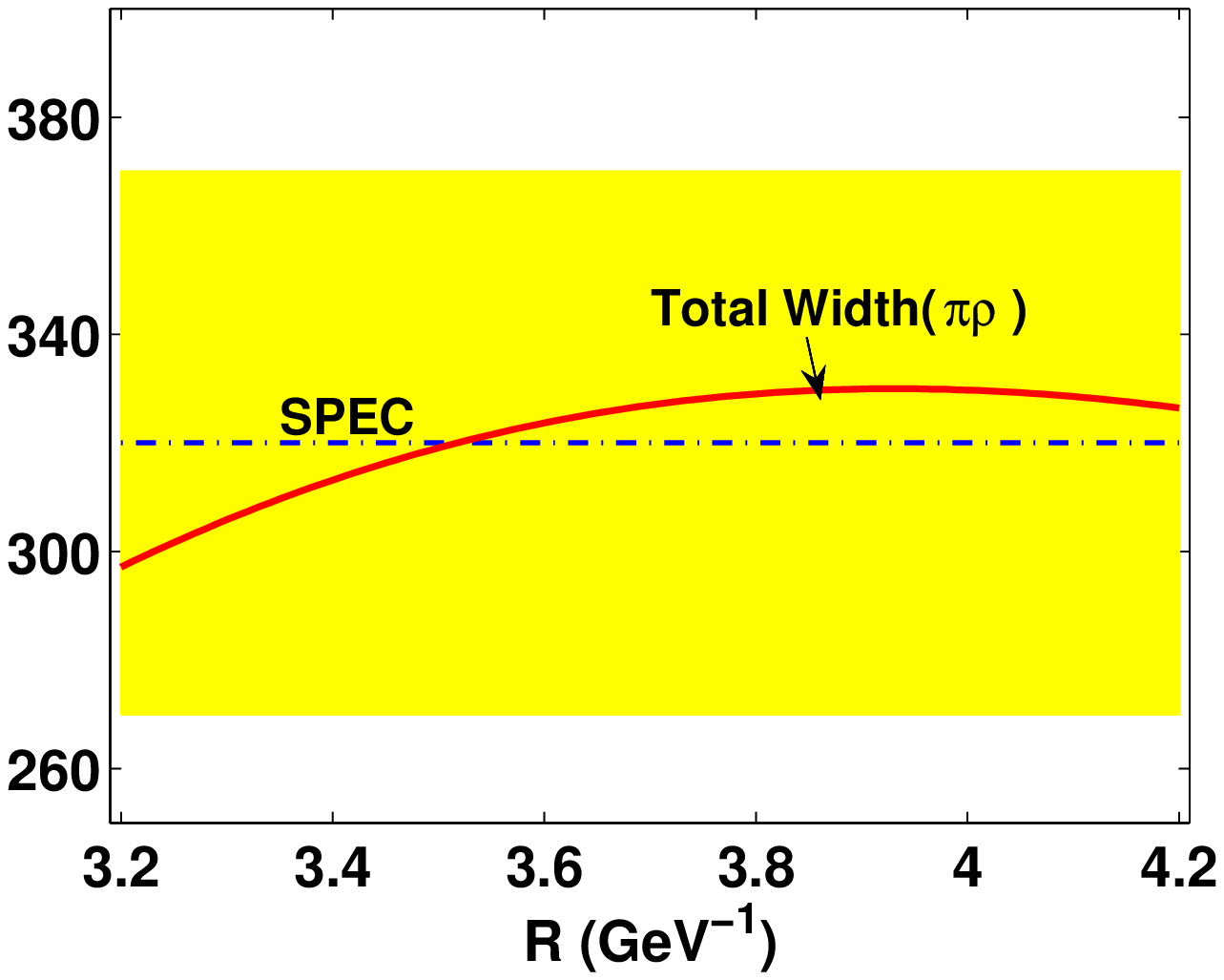}
\includegraphics[scale=0.45]{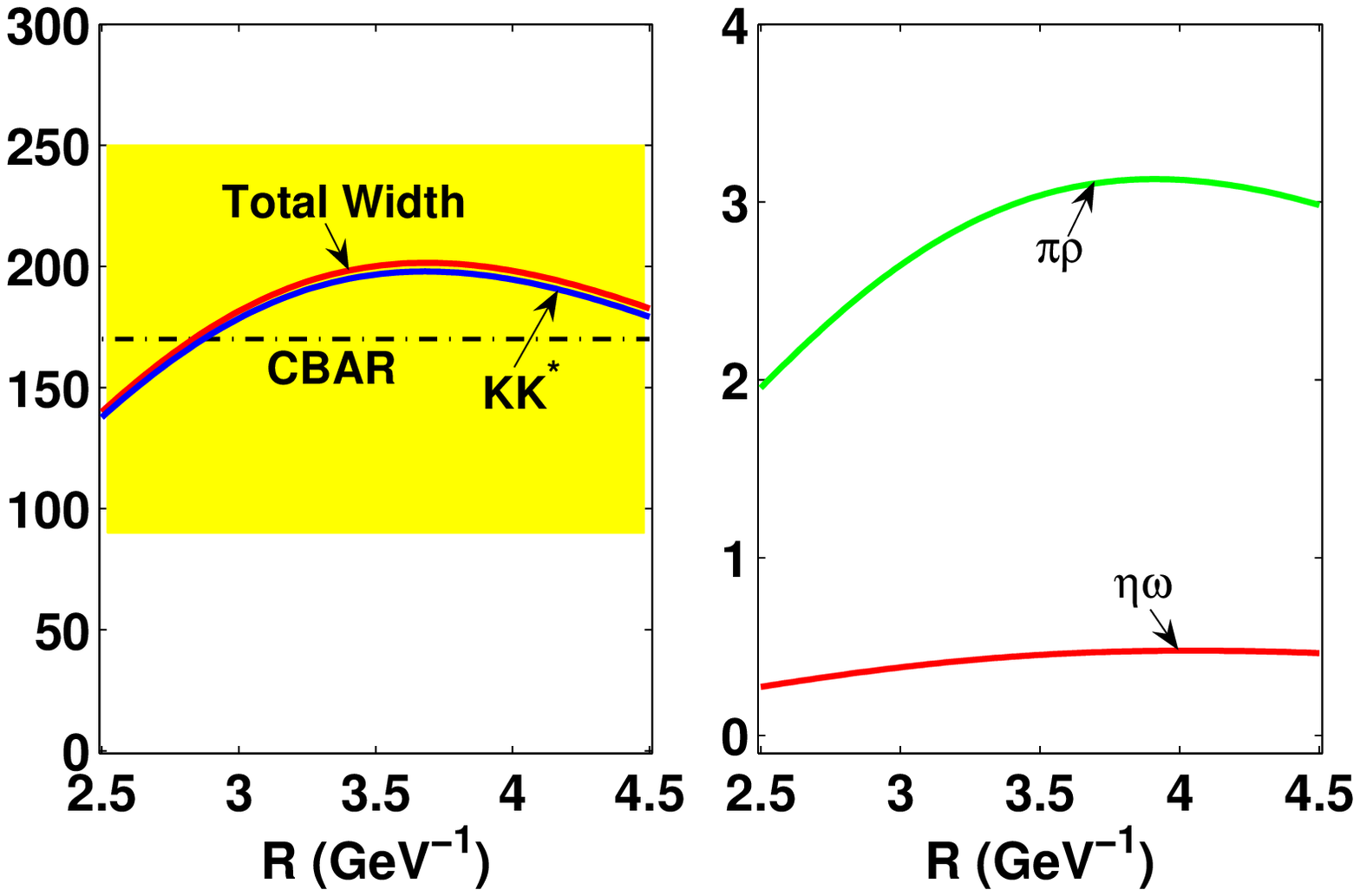}
\caption{(Color online) $R$ dependence of the total decay width of $h_{1}(1170)$ (top) and $h_1(1380)$ (bottom). Here, the dot-dashed lines with the band are
from Refs. \cite{Dankowych:1981ks,Abele:1997vu}. All results are in units of MeV. \label{h11170}}
\end{figure}

The obtained partial and total decay widths of $h_{1}(1170)$ and
$h_1(1380)$ are shown in Fig. \ref{h11170}. Our results indicate that $h_{1}(1170)$ and $h_1(1380)$ are suitable candidates for the ground states in the $h_1$ meson family. Our result that $h_1(1380)$ mainly decays into $KK^{*}$ is consistent
with the experimental fact that $h_{1}(1380)$ has a dominant $s\bar{s}$ component \cite{Abele:1997vu, Aston:1987ak}.

According to the Regge trajectory analysis in Fig. \ref{Regge}, $h_{1}(1595)$, $h_{1}(1965)$, and $h_{1}(2215)$ are the first, second and third radial excitations of $h_1(1170)$. Here, $h_{1}(1595)$,
$h_{1}(1965)$, and $h_{1}(2215)$ have the same flavor wave functions as $h_1(1170)$ in Eq. (\ref{h1eq1}). The mixing angle $\theta_1$ in Eq.
(\ref{h1eq1}) is replaced by
$\theta_2$, $\theta_3$, and $\theta_4$ for  the corresponding $h_1$ states.
As for these higher radial excitations, the mixing
angles $\theta_i$ ($i=2,3,4$) were not well determined. Thus, we take a typical mixing angle {$\theta_i=85.6^\circ$} to discuss the decay behaviors
of $h_{1}(1595)$, $h_{1}(1965)$, and $h_{1}(2215)$.


\begin{figure}[htbp]
\includegraphics[scale=0.45]{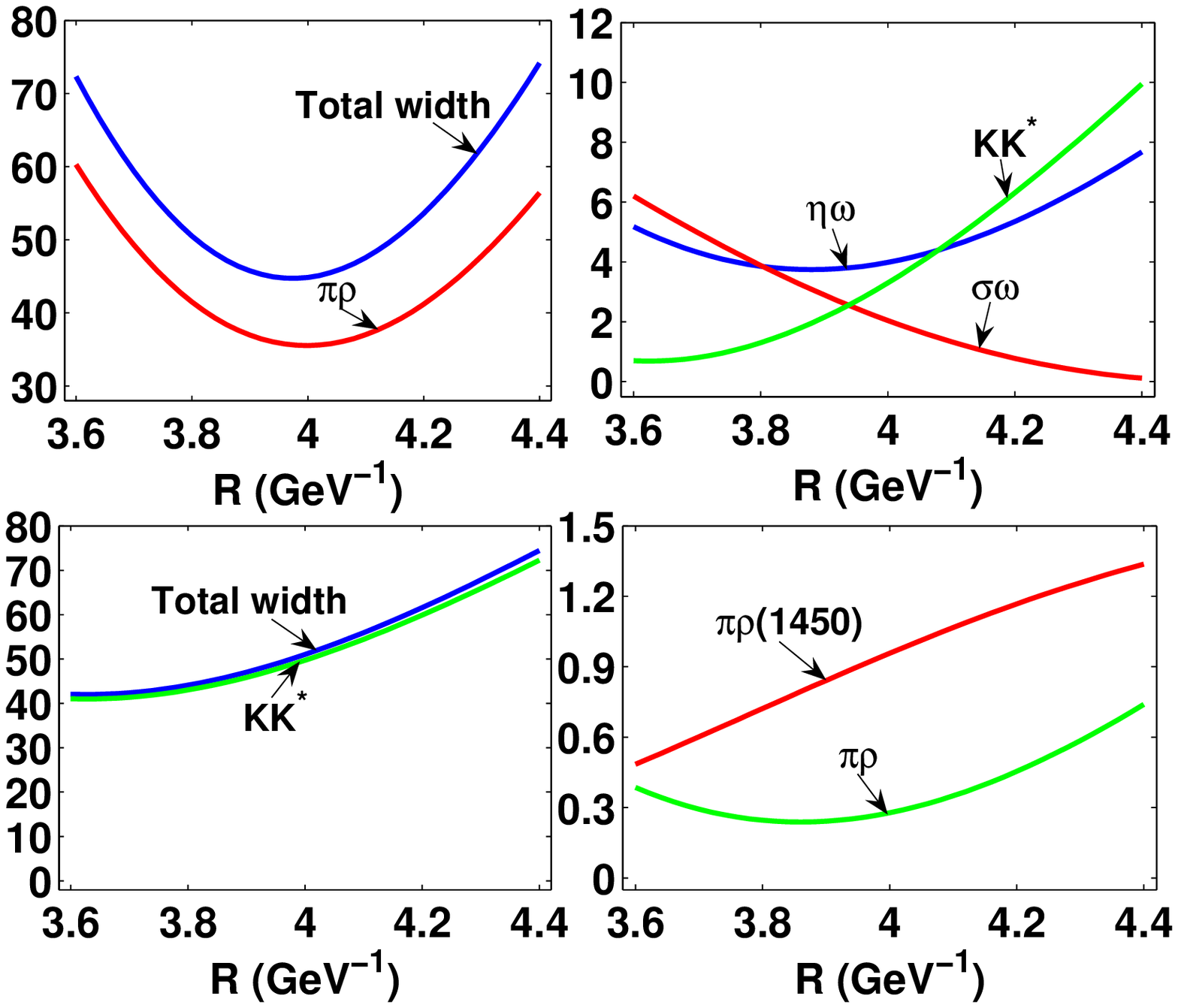}
\caption{(Color online) $R$ dependence of the total and partial decay widths of $h_{1}(1595)$ (top) and $h_{1}(1780)$ (bottom). All results
are in units of MeV. \label{hnewn2}}
\end{figure}

As for $h_{1}(1595)$, we find that the obtained total decay width is much smaller than $384\pm60^{+70}_{-100}$ MeV measured by the BNL-E852 Collaboration
\cite{Eugenio:2000rf}. Thus, we suggest performing a precise measurement of the resonance parameters of $h_{1}(1595)$, which would be helpful in clarifying this discrepancy.
The result shown in Fig.~\ref{hnewn2} indicates that $\pi\rho$ is a dominant decay mode of $h_{1}(1595)$. In addition, $h_{1}(1595)\to \omega\eta$ has a sizable
contribution to the total decay width, which explains why the $\omega\eta$ mode was found in Ref. \cite{Anisovich:2001cr}.
As the predicted partner of $h_{1}(1595)$, $h_{1}(1780)$ dominantly decays into $KK^{*}$, as presented in Fig. \ref{hnewn2}.

\begin{figure}[htbp]
\includegraphics[scale=0.44]{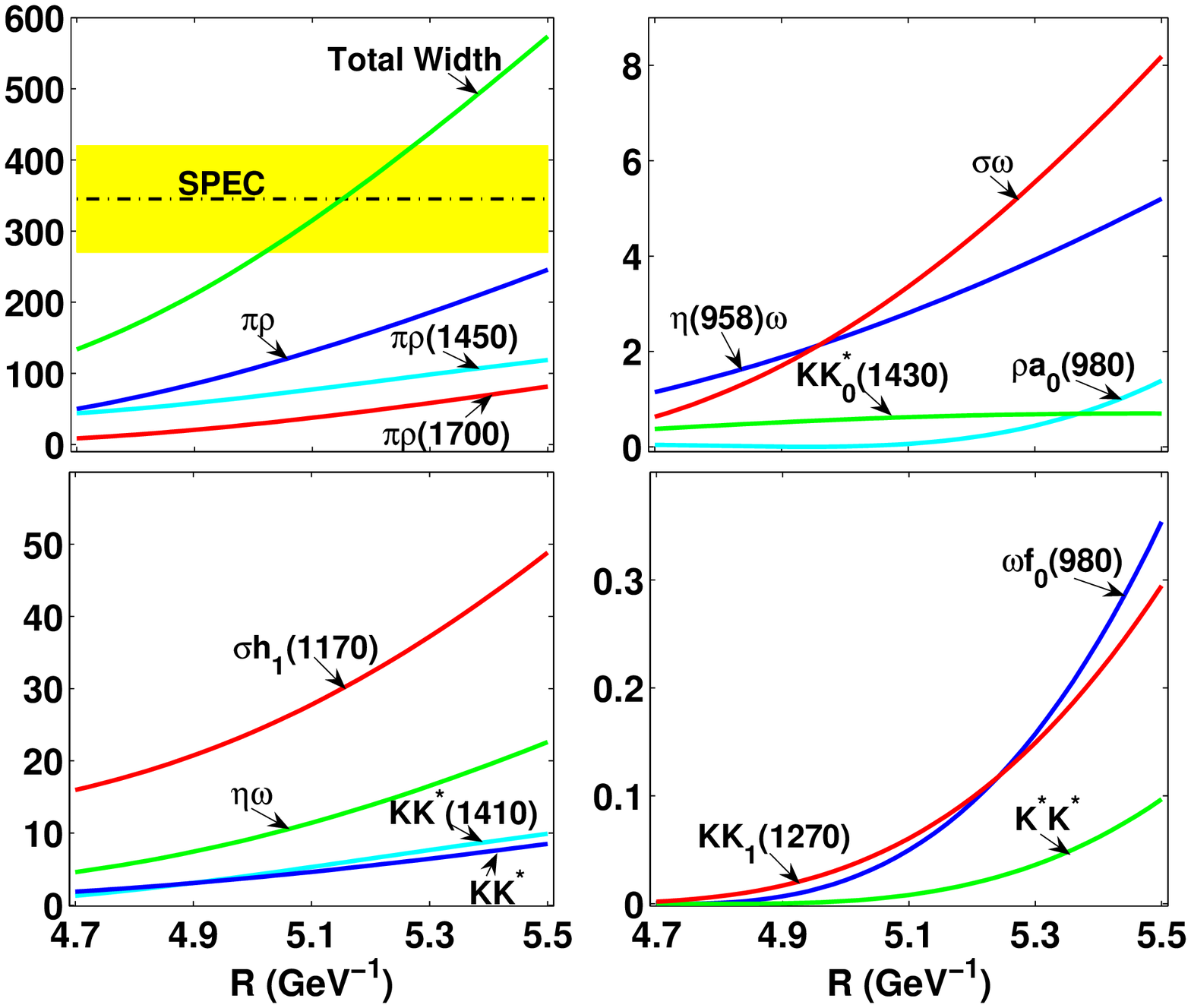}
\caption{(Color online) $R$ dependence of the total and partial decay widths of $h_{1}(1965)$. The experimental total width from Ref.
\cite{Anisovich:2011sva} is denoted by the dot-dashed line with the band. All results are in units of MeV. \label{h11965}}
\end{figure}

\begin{figure}[htbp]
\includegraphics[scale=0.39]{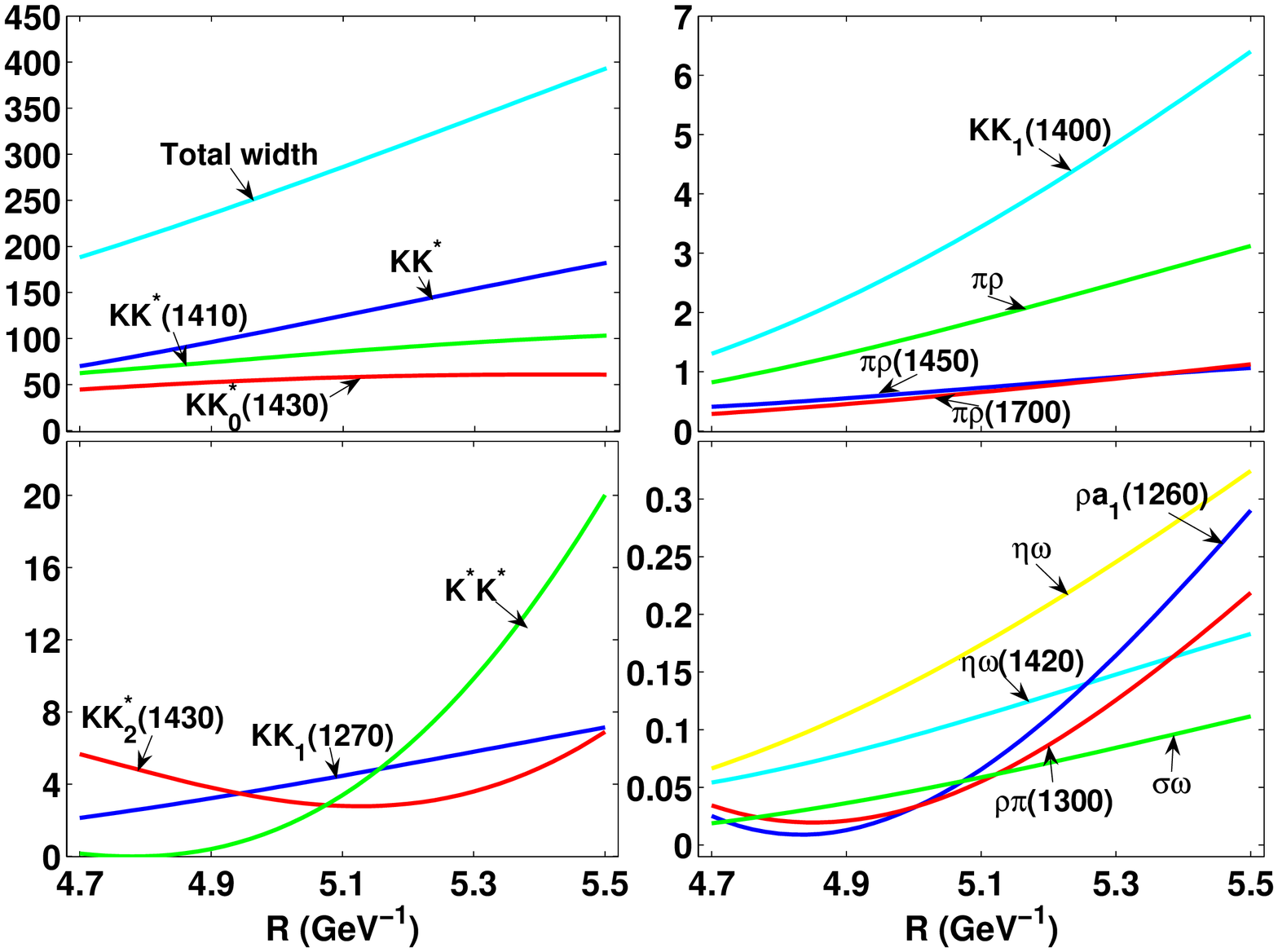}
\caption{(Color online) $R$ dependence of the total and partial decay widths of $h_{1}(2120)$. All results are in units of MeV.
\label{h12120}}
\end{figure}

Figure \ref{h11965} presents the results for $h_{1}(1965)$, where the calculated total decay width can overlap with the Crystal Barrel data \cite{Anisovich:2011sva}
when $R=(5.02\sim5.28)$ GeV$^{-1}$. Its main decay channels are $\pi\rho$, $\pi\rho(1450)$, and $\pi\rho(1700)$, while $\sigma h_{1}(1170)$ also provides a
considerable value. As a partner of $h_{1}(1965)$, $h_{1}(2120)$ is predicted in this work, where its main decay modes are $KK^{*}$, $KK^{*}(1410)$, and
$KK^{*}_{0}(1430)$ (see Fig. \ref{h12120} for more details on its decay properties).

\begin{figure*}[htbp]
\includegraphics[scale=0.45]{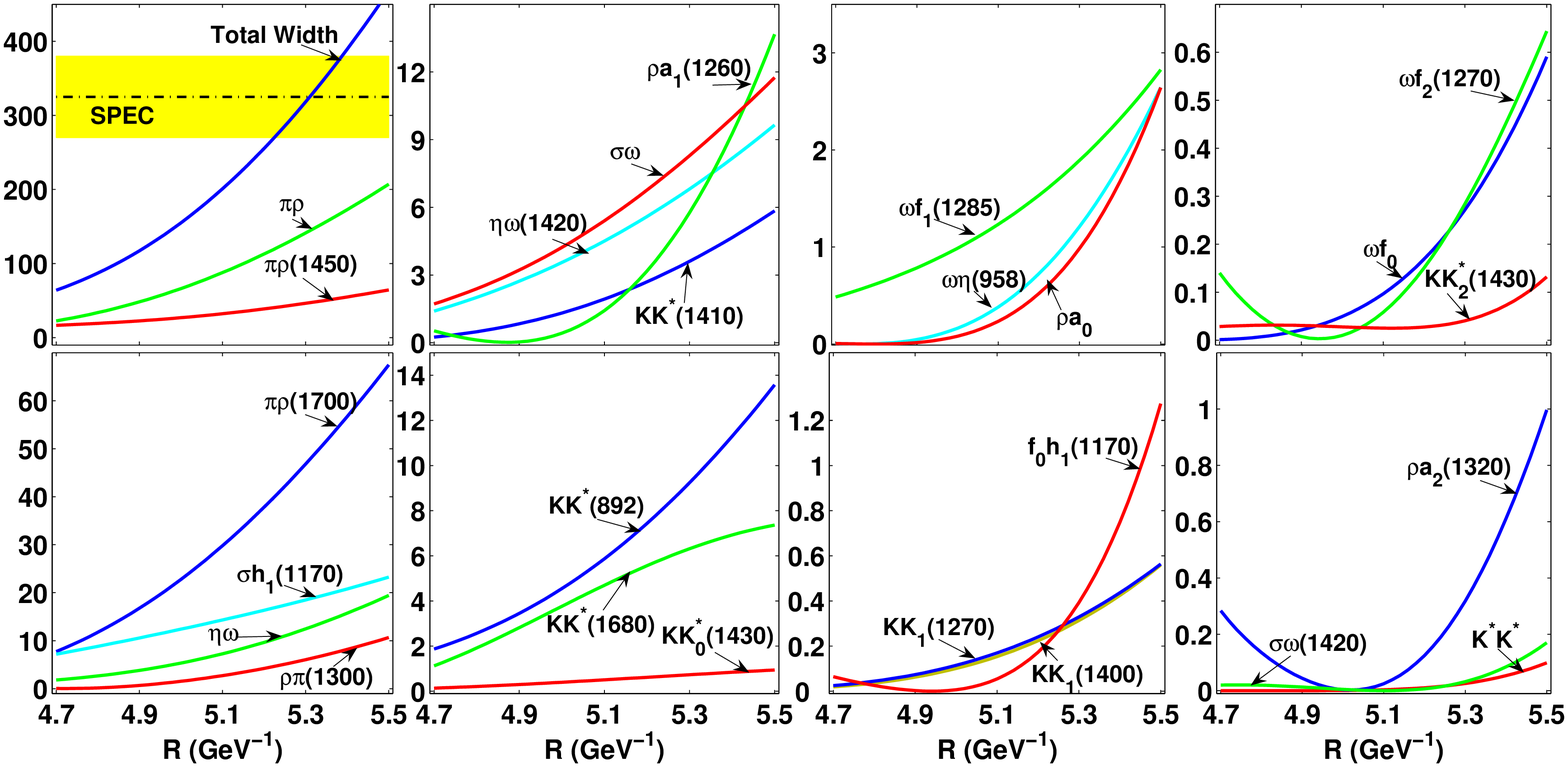}
\caption{(Color online) $R$ dependence of the total decay width of $h_{1}(2215)$. The experimental total width from Ref. \cite{Anisovich:2011sva} is denoted by the
dot-dashed line with the band. All results are in units of MeV. \label{h12215}}
\end{figure*}

\begin{figure}[htbp]
\includegraphics[scale=0.405]{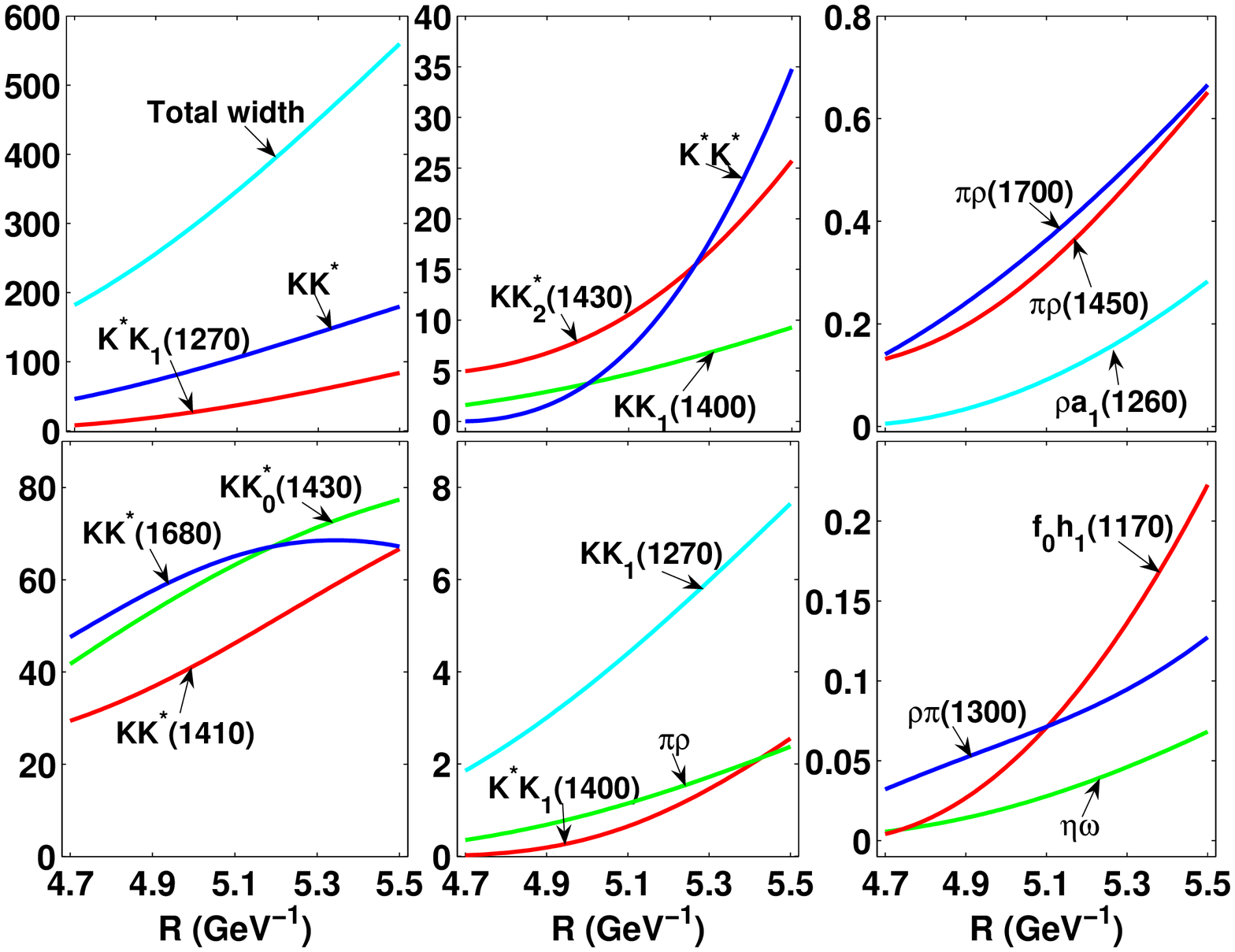}
\caption{(Color online) $R$ dependence of the total decay width of $h_{1}(2340)$. All results are in units of MeV. \label{h12340}}
\end{figure}

The total and partial decay widths of $h_{1}(2215)$ and its partner $h_{1}(2340)$ predicted in this work are listed in Figs. \ref{h12215} and
\ref{h12340}, respectively.
The main decay modes of $h_{1}(2215)$ and $h_{1}(2340)$ can be found in Figs
\ref{h12215} and \ref{h12340}.

\section{Discussion and conclusion}\label{sec3}

Although there are abundant axial-vector states in the PDG \cite{Beringer:1900zz}, the properties of the observed axial-vector states are still unclear.
The present unsatisfactory research status of the observed axial-vector states stimulates us to systematically study them, which will be helpful for
revealing their underlying structures. As a crucial step, we have studied whether the observed axial-vector states can be categorized into the axial-vector meson family.

In this work, we have discussed the observed axial-vector states by assigning them as conventional states in the axial-vector meson family, where both the
analysis of the mass spectra and the calculation of their two-body OZI-allowed strong decays have been performed.

{In our calculation using the QPC model, we took different $R$ ranges to produce the total width of the discussed axial-vector states. For the $a_1$ and $b_1$ states, for example, we
listed the obtained $R$ values for different states (see Table \ref{ab} for more details). We found that
the corresponding $R$ values become larger with an increase in the radial quantum number, which is consistent with our understanding, i.e., the size of higher radial excitations is larger than that of lower radial excitations. Thus, our calculation can reflect this phenomenon, which provides a test of the reliability of our calculation. In addition, we also noticed that states with the same radial quantum number in the $a_1$ and $b_1$ families have similar $R$ ranges, which reflects the fact that the $a_1$ state is the isospin partner of the corresponding $b_1$ state.}

{
When we discussed the decay behaviors of higher radial excitations in the $f_1$ and $h_1$ meson families, we fixed the corresponding mixing angles to match the numerical results, which is due to the absence of a
theoretical study of these mixing angles, and these mixing angles cannot be determined by the present experimental data \cite{Beringer:1900zz}. However, for the ground states of $f_1$ and $h_1$, the situation is totally different, where
the corresponding mixing angles are fixed by experimental data. Thus, in this work we adopted a very simple and crude approach, i.e., we took the same value of the mixing angle for the ground and the corresponding radial excitations.  We expect more experimental data on radial excitations in the $f_1$ and $h_1$ meson families. Then we can carry out further theoretical studies by considering the effect of the mixing angles. }

{
\begin{table}[htbp]
 \renewcommand{\arraystretch}{1.8}
\caption{The obtained $R$ value for these discussed $a_{1}$ and $b_{1}$ states in this work.
 \label{ab}}
\begin{tabular}{lcc|cccccc}
\toprule[1pt]
state&$n^{2S+1}L_{J}$&$R$ (GeV$^{-1}$)& state & $n^{2S+1}L_{J}$&$R$ (GeV$^{-1}$)\\
 \midrule[1pt]
  $a_{1}(1260)$ & $1^{3}P_{1}$ & $3.846$        &$b_{1}(1235)$ &$1^{1}P_{1}$&$3.704$          \\
  $a_{1}(1640)$ & $2^{3}P_{1}$ & $4.30\sim4.64$ &$b_{1}(1640)$ &$2^{1}P_{1}$&                 \\
  $a_{1}(1930)$ & $3^{3}P_{1}$ & $4.58\sim4.92$ &$b_{1}(1960)$ &$3^{1}P_{1}$&$4.66\sim5.16$   \\
  $a_{1}(2095)$ & $3^{3}P_{1}$ & $4.78\sim5.16$ &              &             &                 \\
  $a_{1}(2270)$ & $4^{3}P_{1}$ & $5.12\sim5.32$ &$b_{1}(2240)$ &$4^{1}P_{1}$&$5.20\sim5.54$   \\
\bottomrule[1pt]
\end{tabular}
\end{table}
}

In summary, this phenomenological analysis not only tested possible assignments of the axial-vector states, but also predicted abundant information about their partial
decays, which is valuable for further experimental studies of the observed states. In addition, we have also predicted some missing axial-vector mesons, and have roughly determined their mass values and decay behaviors. We have also suggested an experimental search for the missing states; the BESIII and COMPASS experiments
will be good platforms to carry out the study of light hadron spectra.

\section*{Acknowledgments}

This project is supported by the National Science Foundation for Fostering Talents in Basic Research of the National Natural Science Foundation of China and  the National Natural Science
Foundation of China under Grants No. 11222547 and No. 11175073, the Ministry of Education of China
(SRFDP under Grant No.
2012021111000), and the Fok Ying Tung Education Foundation
(Grant No. 131006).




\vfil

\end{document}